\documentclass[a4paper,12pt]{article}
\usepackage[margin=1.3cm]{geometry}
\usepackage{comment}
\usepackage{amssymb,extarrows,graphicx,subfigure,setspace}
\usepackage{cite}
\usepackage{slashed}
\usepackage{tensor}
\usepackage[toc,page]{appendix}
\usepackage{color}
\usepackage{physics}
\usepackage{hyperref}
\usepackage{dirtytalk}
\hypersetup{colorlinks=true, linkcolor=blue, citecolor=red, linktoc=page}
\makeatother

\usepackage{float}
\usepackage{textcomp}
\usepackage{amsmath}
\newmuskip\pFqmuskip

\newcommand*\pFq[6][8]{%
  \begingroup 
  \pFqmuskip=#1mu\relax
  \mathcode`\,=\string"8000
  \begingroup\lccode`\~=`\,
  \lowercase{\endgroup\let~}\pFqcomma
  {}_{#2}F_{#3}{\left[\genfrac..{0pt}{}{#4}{#5};#6\right]}%
  \endgroup
}
\newcommand{\pFqcomma}{\mskip\pFqmuskip}

\usepackage{mathrsfs}
\usepackage{hyperref}

\newcommand{\be}{\begin{equation}}
\newcommand{\bea}{\begin{eqnarray}}
\newcommand{\eea}{\end{eqnarray}}
\newcommand{\ba}{\begin{array}}
\newcommand{\ea}{\end{array}}
\newcommand{\ee}{\end{equation}}
\newcommand{\bes}{\begin{equation*}}
\newcommand{\beas}{\begin{eqnarray*}}
\newcommand{\eeas}{\end{eqnarray*}}
\newcommand{\bas}{\begin{array*}}
\newcommand{\eas}{\end{array*}}
\newcommand{\ees}{\end{equation*}}

\numberwithin{equation}{section}

\begin{document}
\color{black}
\begin{center}
\Large{\bf Thermodynamic Topology of Kiselev-AdS Black Holes\\ within f (R, T)  gravity }\\
\small \vspace{1cm}
{\bf Saeed Noori Gashti $^{\dag}$\footnote {Email:~~~saeed.noorigashti@stu.umz.ac.ir}},\quad
{\bf Mohammad Ali S. Afshar $^{\star}$\footnote {Email:~~~m.a.s.afshar@gmail.com}},\quad
{\bf Mohammad Reza Alipour$^{\dag,\star}$\footnote {Email:~~~mr.alipour@stu.umz.ac.ir}},\quad\\
\vspace{0.2cm}{\bf Yassine Sekhmani $^{\ddag,\S}$\footnote {Email:~~~sekhmaniyassine@gmail.com}}
{\bf Jafar Sadeghi$^{\star}$\footnote {Email:~~~pouriya@ipm.ir}},\quad
{\bf Javlon Rayimbaev$^{a,b,c,d}$\footnote {Email:~~~javlon@astrin.uz}}\\
\vspace{0.5cm}$^{\star}${Department of Physics, Faculty of Basic
Sciences,\\
University of Mazandaran
P. O. Box 47416-95447, Babolsar, Iran}\\
\vspace{0.3cm}$^{\dag}${School of Physics, Damghan University, P. O. Box 3671641167, Damghan, Iran}\\
\vspace{0.2cm}$^{\ddag}${Center for Theoretical Physics, Khazar University, 41 Mehseti Street, Baku, AZ1096, Azerbaijan}\\
\vspace{0.3cm}$^{\S}${Centre for Research Impact \& Outcome, Chitkara University Institute of Engineering and Technology, Chitkara University, Rajpura, 140401, Punjab, India}\\
\vspace{0.2cm}$^{a}${Institute of Fundamental and Applied Research, National Research University TIIAME,\\ Kori Niyoziy 39, Tashkent 100000, Uzbekistan}\\
\vspace{0.2cm}$^{b}${University of Tashkent for Applied Sciences, Str. Gavhar 1, Tashkent 100149, Uzbekistan}\\
\vspace{0.2cm}$^{c}${Urgench State University, Kh. Alimdjan str. 14, Urgench 220100, Uzbekistan}\\
\vspace{0.2cm}$^{d}${Tashkent State Technical University, Tashkent 100095, Uzbekistan}
\small \vspace{0.5cm}
\end{center}
\begin{abstract}
In this paper, we investigate the topological charge and the conditions for the existence of the photon sphere (PS) in Kiselev-AdS black holes within \(f(R, T)\) gravity. Furthermore, we establish their topological classifications. We employ two different methods based on Duan’s topological current \(\phi\)-mapping theory viz analysis of temperature and the generalized Helmholtz free energy methods to study the topological classes of our black hole. By considering the mentioned black hole, we discuss the critical and zero points (topological charges and topological numbers) for different parameters. Our findings reveal that the Kiselev parameter \(\omega\) and the \(f(R, T)\) gravity parameter \(\gamma\) influence the number of topological charges of black holes, leading to novel insights into topological classifications. We observe that for given values of the free parameters, there exist total topological charges (\(Q_{total} = -1\)) for T-method and total topological numbers (\(W = +1\)) for the generalized Helmholtz free energy method. Our research findings elucidate that, in contrast to the scenario where \(\omega = 1/3\), in other cases, increasing the parameter \(\gamma\) increases the number of total topological charges for the black hole. Interestingly, for the phantom field (\(\omega = -4/3\)), we observed that decreasing the parameter \(\gamma\) increases the number of topological charges.
Additionally, we study the results for the photon sphere. The studied models reveal that the simultaneous presence of \(\gamma\) and \(\omega\) effectively expands the permissible range for \(\gamma\). In other words, the model can exhibit black hole behavior over a larger domain. Additionally, it is evident that with the stepwise reduction of \(\omega\), the region covered by singularity also diminishes and becomes more restricted. However, An interesting point about all three ranges is the elimination of the forbidden region in this model. In other words, it appears that in this model and the investigated areas, there is no region where both the \(\phi\) function and the metric function simultaneously lack solutions. Also, at the end, we fully checked the curvatures singularities, and energy conditions for the mentioned black hole.
\end{abstract}
\tableofcontents
\section{Introduction}
Black hole thermodynamics is a fascinating field that explores the intersection of thermodynamics, quantum mechanics, and general relativity. To obtain insights into quantum gravity, researchers investigate various approaches, including the study of black holes and their thermodynamic properties, which serve as a promising tool \cite{1,2}. This approach combines classical thermodynamics with statistical mechanics to interpret the entropy of black holes as the number of quantum degrees of freedom near the event horizon. The thermodynamics of black holes have been explored within different frameworks, with results documented in \cite{3,4,5,6,7,8,9,10,11,12,13,14,15,16,16',16'',16a}.
A key concept in this field is the Hawking-Page phase transition \cite{17}, which highlights a specific relationship between gravity and thermodynamics. This concept suggests a phase transition between thermal radiation in AdS space and black holes, known as confinement/deconfinement phase transitions in the context of the AdS/CFT correspondence.
Additionally, in the thermodynamics of black holes, the study of phase transitions and critical points consistently garners significant attention\cite{a,b,c,d,e}. Notably, the structural similarities observed between the liquid-to-gas phase transition in van der Waals fluids and the first-order phase transition between small and large black holes are particularly noteworthy.
\\\\
Recently, innovative methods have emerged to analyze and compute critical points and phase transitions in black hole thermodynamics. One notable approach is the topological method, as detailed in \cite{18,19,20,21}. To adopt a topological viewpoint in thermodynamics, Duan’s topological current $\phi$ mapping theory is highly recommended.
SW Wei et al. proposed two distinct methods for studying topological thermodynamics, focusing on temperature and generalized free energy functions\cite{18,19}.
The temperature method involves analyzing the temperature function by removing pressure and utilizing the auxiliary and topological parameter $1/\sin \theta$. Based on these assumptions, the potential is then constructed. The generalized free energy function method treats black holes as defects in the thermodynamic parameter space. Their solutions can be explored using the generalized off-shell free energy. In this framework, the stability and instability of the black hole solutions are indicated by positive and negative winding numbers, respectively. For more study, see \cite{18,19,20,21,22,23,24,25,26,27,28,29,30,31,32,33,34,35,36,37,38,39,40,41,42,43,44,44a,44b,44c,44d}.\\
Researchers, by utilizing black hole actions and extracting their effective potential, have discovered that the maximum and minimum of the potential function, which indicate changes in the field structure, can be examined by transforming the potential into a vector field and mapping it onto a two-dimensional space in ($r , \Theta$) coordinates. In this scenario, these maximum and minimum, similar to winding numbers, manifest as rotations around zero points or poles. Since the mathematical structure of this method was designed for scalars, SW Wei extended it to thermodynamics and used it to study temperature, meaning that the coordinates of the location of temperature changes could also play the role of the pole or zero point for a vector field made of temperature\cite{18,19,4000,4001,4002,4003}. In the T method, the goal is to obtain critical points and determine their position using the topological charge of the system. In this method, we do not know much about the phase transition and how to do it, which makes it a simpler method. Then SW Wei and his colleagues extended this method to a more comprehensive potential function, namely the internal energy or, more specifically, the Helmholtz free energy. In the F method, we use the Helmholtz free energy the structure of the $\tau$ function, and the topological charge to determine which phase transition the structure undergoes. If the process has certain maxima and minima in a radial interval, then the system will undergo a first-order phase transition with the creation of an intermediate black hole, and we will observe the process of creation (maximum) and annihilation (minimum) during the process. However, if the function does not have a maximum or minimum and undergoes a uniform temperature trend during the radial changes, the small black hole will directly and gradually turn into a large black hole, and we will witness a second-order phase transition. Both scenarios can be identified using the F method\cite{18,19,4000,4001,4002,4003}. In other words, the F method and the T method are two distinct approaches used to study the thermodynamic topology of black holes. The generalized free energy method involves analyzing the free energy landscape of black holes. By examining the free energy, researchers can identify critical points and phase transitions. The stability of black hole solutions is determined by the sign of the second derivative of the free energy. Positive values indicate stability, while negative values indicate instability. The method uses the concept of topological charge, which is related to the winding numbers of the free energy function. These winding numbers help classify black holes based on their topological properties\cite{18,19,4000,4001,4002,4003}. T Method focuses on the temperature function of black holes. It involves studying the temperature function by removing pressure and introducing an auxiliary topological parameter. Similar to the F method, the T method also considers topological aspects, but it does so through the lens of temperature variations and their impact on the system.  Both methods provide a topological perspective on black hole thermodynamics, allowing researchers to classify black holes based on their topological properties. Both methods aim to determine the stability and instability of black hole solutions, although they use different functions (free energy vs. temperature) to achieve this. While the F method focuses on free energy and the T method on temperature, they are complementary approaches that together offer a more comprehensive understanding of black hole thermodynamics. By using both the F method and the T method, researchers can gain a deeper and more nuanced understanding of the thermodynamic properties and stability of black holes. Each method provides unique insights that, when combined, enhance our overall knowledge of black hole thermodynamics\cite{18,19,4000,4001,4002,4003}.\\
The concept of the topological photon sphere in black holes is a fascinating area of study in theoretical physics. The photon sphere is a region where gravity is so strong that photons (light particles) are forced to travel in orbits. This sphere is crucial for understanding the properties of black holes, including their shadow and the bending of light around them\cite{45,46}. In the context of topology, researchers explore the properties and behaviors of these photon spheres using topological methods. These methods can provide insights into the stability and structure of the photon sphere, as well as its interactions with other physical phenomena in the black hole's vicinity.\\\\The \( f(R, T) \) gravity model presents an intriguing extension of Einstein's equations of motion by incorporating terms related to the Ricci scalar and the trace of the energy-momentum tensor. This theoretical framework has recently attracted considerable attention due to its potential to explain various cosmological phenomena. Additionally, it offers an alternative gravity theory that can account for dark matter and dark energy. Numerous studies have been conducted in the context of \( f(R, T) \) gravity, as noted in the literature\cite{46a,46b,46c,46d,46e,46f,46g,46h,46i,46j,46k,46l,46m}.
According to general relativity (GR), black holes possess gravitational forces so intense that nothing, not even light, can escape their pull. It is fascinating to consider that the effects of \( f(R, T) \) gravity might be more significant for compact objects like black holes. Therefore, exploring black holes within the \( f(R, T) \) gravity framework is particularly compelling. Recognizing this, Santos et al. formulated a black hole solution similar to Kiselev's within the \( f(R, T) \) gravity framework\cite{46n}. Subsequently, the rotating version of this black hole solution has been further elaborated upon\cite{47}. \\Kiselev-AdS black holes within the \( f(R, T) \) gravity framework are an intriguing area of study in theoretical physics. These black holes are solutions to the modified Einstein field equations, incorporating both the Ricci scalar \( R \) and the trace of the energy-momentum tensor \( T \). This framework allows for the exploration of various cosmological and astrophysical phenomena, including the effects of dark energy and dark matter\cite{47,48,49}.
Recent research has focused on constructing rotating Kiselev black holes within this framework using the revised Newman-Janis algorithm. These solutions include special cases like the Kerr and Kerr-Newman black holes and exhibit unique spacetime structures influenced by the parameters of the \( f(R, T) \) function.
Additionally, studies have shown that Kiselev-AdS black holes in \( f(R, T) \) gravity can exhibit phase transitions similar to those seen in thermodynamic systems, such as van der Waals-like and Hawking-Page-like transitions. These phase transitions are influenced by the state parameter \( \omega \) and the \( f(R, T) \) gravity parameter \( \gamma \), providing insights into the nature of dark energy and the thermodynamic properties of black holes\cite{47,48,49}\\\\
These concepts motivated us to study the Kiselev-AdS black holes from \( f(R, T) \) gravity based on thermodynamic topology. We organize the paper as follows:
In Section 2, we briefly introduce the mentioned black hole. Also, we study the curvature singularities and energy conditions of Kiselev-AdS black holes from \( f(R, T) \) gravity in Section 3. We investigate the thermodynamic topology from the perspective of topology using the generalized off-shell Helmholtz free energy and temperature methods and study the photon sphere of Kiselev-AdS black holes from \( f(R, T) \) gravity in Section 4. Finally, we describe the results of our work in detail in Section 5.
\section{Kiselev-AdS black holes from \( f(R, T) \) gravity}
In this study, we aim to obtain a static spherically symmetric AdS black hole solution as proposed by Kiselev within the framework of \( f(R, T) \) gravity. In our analysis, we consider the specific form \( f(R, T) = R + 2 f(T) \)\cite{48,53}. One can assume a static geometry in a spherically symmetric space-time, described by the following line element,
\begin{equation}\label{1}
ds^2=B(r)dt^2-A(r)dr^2-r^2(d\theta^2+\sin^2\theta d\phi^2).
\end{equation}
With respect to \cite{48} and using the form \( f(T) = \gamma T \), the unknown functions is obtained as follows,
\begin{equation}\label{200}
B(r)=\frac{1}{A(r)}=1-\frac{2M}{r}+\frac{r^2}{\ell^2}+\frac{k}{r^{\frac{8(3\pi\omega+\omega\gamma+\pi)}{3\gamma+8\pi-\omega\gamma}}}
\end{equation}
Here, \(\gamma\) is the \( f(R, T) \) gravity parameter, \(M\) is the mass of the black hole, and \(k\) is a constant of integration. By setting \(\gamma \rightarrow 0\), the well-known Kiselev-AdS black hole solution is recovered as follows in General Relativity \cite{54},
\begin{equation}\label{3}
B(r)|_{\gamma\rightarrow0}=1-\frac{2M}{r}+\frac{k}{r^{(3\omega+1)}}+\frac{r^2}{\ell^2}
\end{equation}
The black hole mass can be determined by solving \( B(r_+) = 0 \) as follows,
\begin{equation}\label{4}
M=\frac{1}{2} r_+ \left(k r_+^{-\frac{8 \gamma  \omega +24 \pi  \omega +8 \pi }{-\gamma\omega +3 \gamma +8 \pi }}+\frac{r_+^2}{\ell^2}+1\right)
\end{equation}
The Hawking temperature defined as \( T = \kappa / 2\pi \). So, it can be expressed as follows,
\begin{equation}\label{5}
T=\frac{1}{4 \pi }\bigg(\frac{3 k r_+^{-\frac{\gamma  (7 \omega +3)+8 \pi  (3 \omega +2)}{-8 \pi+\gamma  (\omega -3)}} (-8 \pi  \omega +\gamma -3 \omega )}{8 \pi -\gamma  (\omega -3)}+\frac{3 r_+}{l^2}+\frac{1}{r_+}\bigg)
\end{equation}
We use the well-known Hawking-Bekenstein area law to express the entropy \( S_+ \) in terms of the event horizon radius \( r_+ \) as follows:
\begin{equation}\label{6}
S_+=\pi r_+^2
\end{equation}
\section{Curvatures singularities and energy conditions}
To provide a suitable proof of the singularity and uniqueness of the black hole solution, we have to perform some analyses in terms of scalar invariants such as the Ricci scalar, the Ricci squared scalar, and the Kretschmann scalar. The Ricci scalar with respect to the relevant metric can be given as,
\begin{equation}\label{R}
\mathcal{R}=-\frac{6 k (\gamma  (3 \omega -1)+8 \pi  \omega ) [\gamma  (5 \omega -3)+4 \pi  (3 \omega -1)] r^{-\frac{6 (\gamma +4 \pi ) (\omega +1)}{8 \pi -\gamma  (\omega -3)}}}{(\gamma
(\omega -3)-8 \pi )^2}-\frac{12}{\ell^2}.
\end{equation}
The Ricci squared function, on the other hand, can be expressed in relation to the metric function as,
\begin{equation}\label{RR}
\begin{split}
&\mathcal{R}_{\mu\nu}\mathcal{R}^{\mu\nu}= k^2 (-3 \gamma  \omega +\gamma -8 \pi  \omega )^2 \big[\gamma ^2 (\omega  (17 \omega -6)+9)+16 \pi  \gamma  (6 \omega ^2+\omega +3)\\&+16 \pi ^2
   (9 \omega ^2+6 \omega +5)\big] r^{-\frac{12 (\gamma +4 \pi ) (\omega +1)}{8 \pi -\gamma  (\omega -3)}}\bigg/(\gamma  (\omega -3)-8 \pi )^4{}\\
   &+2 k (\gamma  (3
   \omega -1)+8 \pi  \omega ) (\gamma  (5 \omega -3)+4 \pi  (3 \omega -1)) r^{-\frac{6 (\gamma +4 \pi ) (\omega +1)}{8 \pi -\gamma  (\omega -3)}}\bigg/\ell^2 (\gamma  (\omega -3)-8
   \pi )^2{}+\frac{2}{\ell^4}.
\end{split}
\end{equation}
The Kretschmann scalar can be represented as,
\begin{equation}\label{RRR}
\begin{split}
&\mathcal{R}_{\mu\nu\alpha\beta}\mathcal{R}^{\mu\nu\alpha\beta}=\frac{4 (k r^{\frac{8 (\gamma  \omega +3 \pi  \omega +\pi )}{\gamma  (\omega -3)-8 \pi }}+\frac{r^2}{\ell^2}-\frac{2 M}{r})^2}{r^4}+\frac{4 (-\frac{8 k (\gamma
   \omega +3 \pi  \omega +\pi ) r^{\frac{8 (\gamma  \omega +3 \pi  \omega +\pi )}{\gamma  (\omega -3)-8 \pi }-1}}{\gamma  (-\omega )+3 \gamma +8 \pi }+\frac{2 r}{\ell^2}+\frac{2
   M}{r^2})^2}{r^2}\\&+\bigg(8 k (\gamma  \omega +3 \pi  \omega +\pi ) (\gamma  (7 \omega +3)+8 \pi  (3 \omega +2)) r^{-\frac{6 (\gamma +4 \pi ) (\omega +1)}{8 \pi
   -\gamma  (\omega -3)}}\bigg/(\gamma  (\omega -3)-8 \pi )^2+\frac{2}{\ell^2}-\frac{4 M}{r^3}\bigg)^2
   \end{split}
\end{equation}
Upon deeper inspection of the expressions $\eqref{R}$, $\eqref{RR}$ and $\eqref{RRR}$, it can be seen that the black hole metric is singular for all admissible values of the parameters $\omega$ and $\gamma$. Indeed, the existence of the singularity results from the mass term and the relevant term of the constant $k$ in the black hole metric. By constraining $\omega <-1/3$ and $\gamma<0$, the singularity will disappear due to these imposing conditions. Even so, to rule out the singularity made from the mass and charge terms, it may be worth outlining a routine with a non-linear charge distribution function analogous to Ref\cite{Balart:2014cga}. Along this work, we will not consider such a situation and stick to the metric function \eqref{23} for the rest of the analysis. A few remarks about the scalars mentioned show that the Ricci scalar and the Ricci squared are not a function of the black hole mass $M$. On the other hand, the Kretschmann scalars are functions of the black hole mass $M$, so any variation in the black hole mass, Kiselev $f (R, T)$ parameters can bring about substantial variations in these same scalars. Hence, the scalars under consideration show that the black hole solution is indeed unique and that the AdS background, as well as the Kislev structure, significantly modify the black hole spacetime. So, an analysis of scalar invariant limits at $r = 0$ might provide as follows,
\begin{equation}
   \lim\limits_{r\to 0}
    \mathcal{R},\,\mathcal{R}_{\mu\nu}\mathcal{R}^{\mu\nu}, \,
R_{\alpha\beta\mu\nu}R^{\alpha\beta\mu\nu}\approx\infty.
    \end{equation}
    \begin{figure}[h!]
 \begin{center}
 \subfigure[]{
 \includegraphics[height=5.2cm,width=5.2cm]{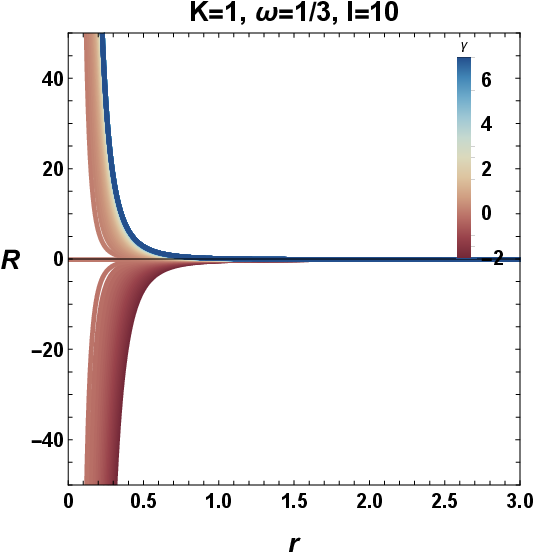}
 \label{r1}}
 \subfigure[]{
 \includegraphics[height=5.2cm,width=5.2cm]{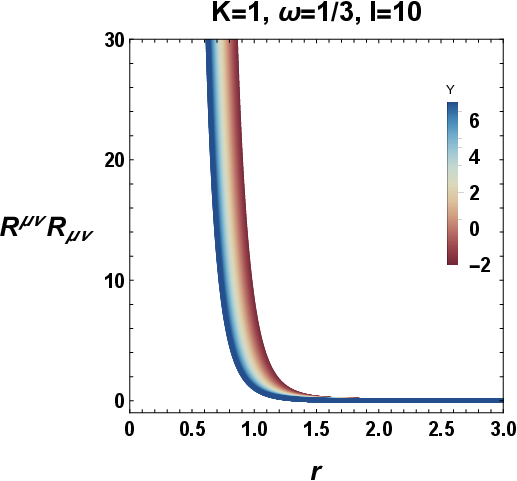}
 \label{r3}}
 \subfigure[]{
 \includegraphics[height=5.2cm,width=5.2cm]{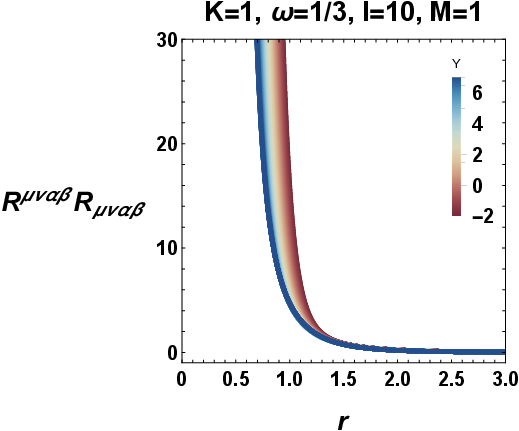}
 \label{r5}}
  \caption{\small{Ricci, Ricci squared and Kretschmann scalars representation for various values of black hole system.}}
 \label{CS}
 \end{center}
 \end{figure}
On the other hand, analyzing at a large distance, the relevant behavior of the scalar invariants can be provided as follows,
\begin{align}
   \lim\limits_{r\to \infty} \begin{cases}
    \mathcal{R}\approx -\frac{12}{\ell^2}
     \\ \mathcal{R}_{\mu\nu}\mathcal{R}^{\mu\nu}\approx\frac{36}{\ell^4}
    \\
R_{\alpha\beta\mu\nu}R^{\alpha\beta\mu\nu}\approx\frac{24}{\ell^4}\nonumber
    \end{cases}\,.
\end{align}
This in turn shows that the Ricci scalar, the Ricci squared scalar, and the Kretschmann scalar comprise a finite term at large distances. In a nutshell, these scalars are proof that our black hole solution is completely consistent and that the parameters of Kiselev, as well as the $f(R, T)$ parameter theory, do not affect the asymptotic behavior of spacetime. The above analysis can be demonstrated based on the three scalar invariants graph. For that reason, Figure \ref{CS} displays the radial variations of scalar invariants in spacetime (\ref{1}), taking into account the Ricci scalar (Fig. \ref{r1}), the square of the Ricci tensor (Fig. \ref{r3}), and the Kretschmann scalar (Fig. \ref{r5}) for various values of the parameter $\gamma$ with $\omega=1/3$. The figure reveals that the scalar invariants show a declining trend along the radial coordinates. Furthermore, as the value of  $\omega$ and $\gamma$ decrease, the Ricci squared scalar and the Kretschmann scalar increase. On the other hand, a proportional relation is observed between the variation of the Ricci scalar and the parameter $\gamma$. Similarly, it is found that all three scalars drop to zero when $r$ is outside the event horizon. We now look at the weak, strong, and dominant energy conditions for the source fluid. The energy-momentum tensor for the anisotropic source fluid is given by \cite{Rajagopal:2014ewa,Delsate:2014zma,Setare:2015xaa,Capozziello:2022ygp}
\begin{equation}\label{aa}
    T^{\mu\nu}=\rho e_0^\mu e_0^\nu+\sum_{i=1}^3P_ie_i^\mu e_i^\nu
\end{equation}
where $\rho$ is the energy density, $P_i\quad (i = 1, 2, 3)$ are the pressures for the source fluid, and $e_i^\mu$ represents the vielbein components. Eq. \ref{aa} refers to the source of the Einstein field equations that might be needed to derive a black hole solution. We consider the static and spherically symmetric line element proposed in Eq. \ref{1}, and we require that this metric satisfies the Einstein equations according to the auxiliary term in Eq. \ref{200}. To be more specific, because we are looking for an Anti-de Sitter-type contribution with  a negative cosmological constant, we obtain that the Einstein field equations give\cite{Rajagopal:2014ewa,Delsate:2014zma,Setare:2015xaa,Capozziello:2022ygp}
\begin{align}
    \rho&=-P_r=\frac{1-B(r)-rB'(r)}{r^2}+8\pi P_\Lambda,\label{ee1}\\
    P_\theta&=P_\phi=\frac{rB''(r)+2B'(r)}{2 r}-8\pi P_\Lambda.\label{ee2}
\end{align}
Notice that we employ the convention $G = 1/(8\pi)$, whereas \cite{Rajagopal:2014ewa} uses $G = 1$, $G$ being Newton's constant. These results are broadly in agreement with the ones given in \cite{Rajagopal:2014ewa, Rajagopal:2014ewa,Delsate:2014zma,Setare:2015xaa,Capozziello:2022ygp}, where $P_r$ is the longitudinal pressure, while $P_\theta$ and $P_\phi$ are the transverse pressures. Moreover, the term $8\pi P_\Lambda$, which is included in equations (\ref{ee1}) and (\ref{ee2}), is essential to ensure the presence of the negative cosmological constant in the Einstein field equations and according to current studies \cite{Gunasekaran:2012dq,Kubiznak:2012wp}, one can interpret it as a thermodynamic pressure, such that $P_\Lambda=-\Lambda/8\pi=3/8\pi \ell^2$. We will now examine the energy conditions in question for the black hole solution~\cite{Hawking:1973uf, Ghosh:2008zza,Kothawala:2004fy, Ghosh:2020syx}.
  \begin{figure}[h!]
 \begin{center}
 \subfigure[]{
 \includegraphics[height=5.2cm,width=5.2cm]{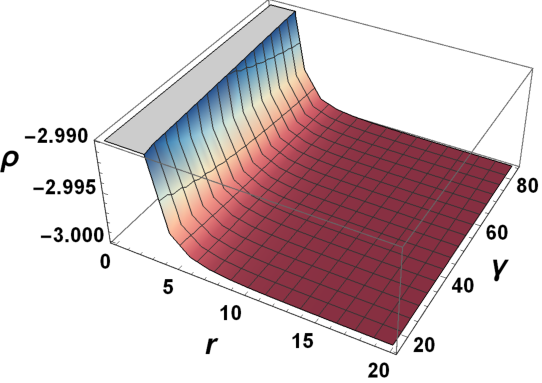}
 \label{wec}}
 \subfigure[]{
 \includegraphics[height=5.2cm,width=5.2cm]{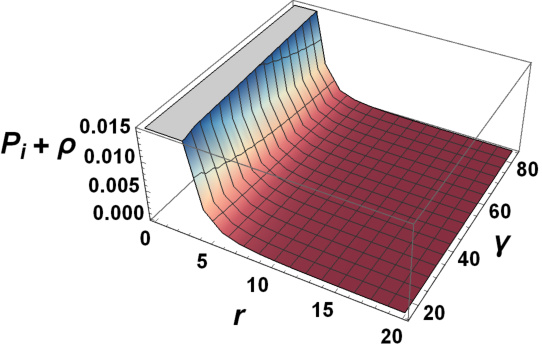}
 \label{nec}}
 \subfigure[]{
 \includegraphics[height=5.2cm,width=5.2cm]{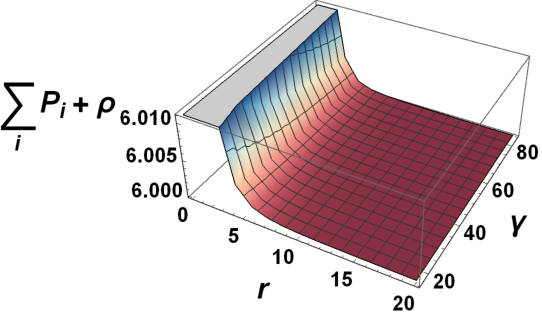}
 \label{sec}}
  \caption{\small{Energy conditions using $\omega=1/3$ and $k=1$.}}
 \label{fig2}
 \end{center}
 \end{figure}
The elements of the stress-energy tensor $T^{\mu\nu}$ for Kiselev-AdS black hole within $f(R, T)$ gravity are as follows,
        \begin{align}
        \rho&=\frac{3 k (\gamma  (3 \omega -1)+8 \pi  \omega ) r^{-\frac{6 (\gamma +4 \pi ) (\omega +1)}{8 \pi -\gamma  (\omega -3)}}}{8 \pi -\gamma  (\omega -3)} =-P_r\label{rho}\\
      P_\theta&=P_\phi=3 \left(\frac{4 k (\gamma  \omega +3 \pi  \omega +\pi ) (\gamma  (3 \omega -1)+8 \pi  \omega ) r^{-\frac{6 (\gamma +4 \pi ) (\omega +1)}{8 \pi -\gamma  (\omega -3)}}}{(\gamma
   (\omega -3)-8 \pi )^2}\right).\label{P}
    \end{align}
    \begin{itemize}
        \item For each time vector $t^\mu$, the weak energy condition (WEC) demands that $T_{\mu\nu}\, t^\mu t^\nu\geqslant0$ everywhere. This is equivalent to~\cite{Toshmatov:2017kmw}
    \begin{equation}
        \rho\ge0,\quad \rho+ P_i\ge0\quad (i=r, \theta, \phi)
    \end{equation}
    and so $\rho+P_r=0$ and
    \begin{align}
        &\rho+P_\theta=\frac{9 (\gamma +4 \pi ) k (\omega +1) (\gamma  (3 \omega -1)+8 \pi  \omega ) r^{-\frac{6 (\gamma +4 \pi ) (\omega +1)}{8 \pi -\gamma  (\omega -3)}}}{(\gamma  (\omega -3)-8
   \pi )^2}.
   \label{3-}
    \end{align}
Given that $\omega>1/3$ and that $\gamma>0$, the WEC is satisfied. It is evident from Fig. $\ref{wec}$ that, given a span of horizon radii, the energy density $\rho$ is positive in accordance with the value of fixed $\omega$ and for a spectral value of the parameter $\gamma$.
\item For each null vector $t^\mu$ in the entire spacetime, the null energy condition (NEC) requires that $T_{\mu\nu}\, t^\mu t^\nu\geqslant0$. When $\omega>1/3$, the NEC predicts $\rho+P_r \geqslant 0$, which is identically zero, and $\rho+P_\theta=\rho+P_\phi \geqslant 0$, which is satisfied for equation \eqref{3-}.
\item The strong energy condition (SEC) states that, for every time vector $t^\mu$, $T_{\mu\nu}\, t^\mu t^\nu\geqslant 1/2 \,T_{\mu\nu} t^\nu t_\nu$ globally, assuming that~\cite{Toshmatov:2017kmw}
\begin{equation}
        \rho+\sum_i P_i=P_r+2\,P_\theta\ge0.
        \label{37}
    \end{equation}
   Upon careful inspection, the following restriction is the only one that satisfies the SEC:
     \begin{align}
     \omega>1/3\quad \text{and} \quad \gamma>0
    \end{align}
    \end{itemize}
Graphical analysis shows that energy conditions such as WEC, NEC, and SEC vary as a function of the variable $r$ (see Fig. \ref{fig2}). Thus, we observe that for each choice of values of $\omega$ with a range of values of the parameter $\gamma$, it is observed that the WEC is satisfied for the considered case such as $\omega=1/3$ with $16\leqslant\gamma\leqslant84.5$ (Fig. \ref{wec}). Similarly, the NEC seems to be satisfied also for the case such that $\omega=1/3$ with $16\leqslant\gamma\leqslant84.5$ (Fig. \ref{nec}).
The Kiselev predominantly satisfies the SEC, as shown in Fig. \ref{sec}. This is akin in contrast to the quintessence of dark energy within GR. Indeed, a violation of the SEC is understood as a violation of the attractive pattern of gravity, as exemplified by dark energy accelerating the expansion of the universe in cosmological studies, together with the matter content of the background of a regular black hole, whose singularity has been superseded by a Sitter core. For brevity, it is worth noting that the general mapping between SEC and gravity behavior is the focus of an ongoing search in the literature. It is well known that the validity of the SEC in a gravity environment is typically related to the attractive aspect of gravity. This is true in GR, where the SEC must be assumed to secure the attractive nature of the theory - \emph{focus theorem}~\cite{Carroll:2004st, Wald:1984rg}. Nevertheless, such a connection need not have entirely general validity in extended gravity. A paradigmatic illustration is provided by $f(R)$ gravity, shown in~\cite{Santos:2016vjg}, involving the Raychaudhuri equation. It has already been revealed that even supposing the standard SEC, the Raychaudhuri equation can present positive inputs from spacetime geometry, generally considered a possible sign of repulsive gravity. As a result, the SEC/attractive gravity paradigm appears invalid in this context. A piece of stringent and comprehensive evidence that the attractiveness of gravity is no longer guaranteed by the SEC in extended gravity is beyond the present analysis. This task will be left to the future.
\section{Thermodynamics topology and photon spheres}
The photon sphere is a region surrounding a black hole where the gravitational pull is so intense that it forces photons into circular orbits. This sphere is crucial because it defines the boundary within which light can be trapped in orbit around the black hole, leading to intriguing phenomena such as gravitational lensing and the black hole's shadow\cite{50,51,52,52',52'',52''',52a,52b,4005}. Thermodynamic topology examines the topological characteristics of black holes about their thermodynamic behavior. This method aids in understanding phase transitions and the stability of black holes. By studying the topological charge and critical points, researchers can categorize different phases of black holes and forecast their behavior under various conditions. For example, in hyperscaling-violating black holes, the topological charge can reveal whether the black hole is stable or if it will transition to a different phase. These concepts are essential for comprehending the complex nature of black holes and their interactions with light and matter. In the following section, we will discuss these concepts in the context of Kiselev-AdS black holes within \(f(R, T)\) gravity.
\subsection{Topology of black holes thermodynamics}
\subsubsection{F-Method}
\begin{figure}[h!]
 \begin{center}
 \subfigure[]{
 \includegraphics[height=4.2cm,width=4.2cm]{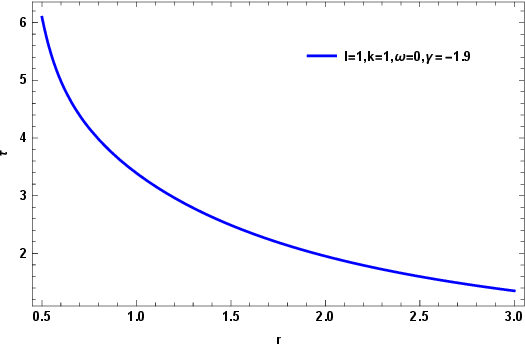}
 \label{1a}}
 \subfigure[]{
 \includegraphics[height=4.2cm,width=4.2cm]{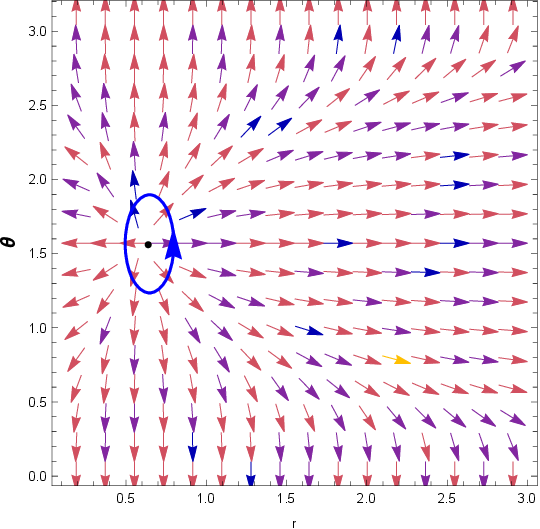}
 \label{1b}}
 \subfigure[]{
 \includegraphics[height=4.2cm,width=4.2cm]{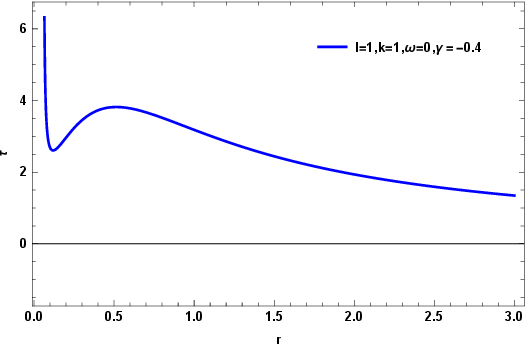}
 \label{1c}}
\subfigure[]{
 \includegraphics[height=4.2cm,width=4.2cm]{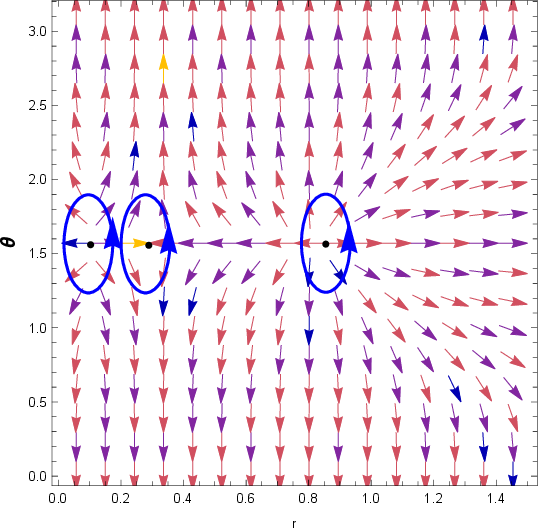}
 \label{1d}}
  \caption{\small{The curve corresponding to Eq.(\ref{16}) is depicted in Figs. \ref{1a} and \ref{1c}. In Figs. \ref{1b} and \ref{1d}, the colorful arrows illustrate the vector field (n) on a segment of the $(r-\theta)$ plane for Kiselev-AdS black holes within f(R, T) gravity, with parameters $(\ell=k=1, \omega=0, \gamma=-1.9)$ and $(\ell=k=1, \omega=0, \gamma=-0.4)$, respectively. The ZPs are positioned at $(r, \theta)$ on the circular loops.}}
 \label{F1}
 \end{center}
 \end{figure}

\begin{figure}[h!]
 \begin{center}
 \subfigure[]{
 \includegraphics[height=4.2cm,width=4.2cm]{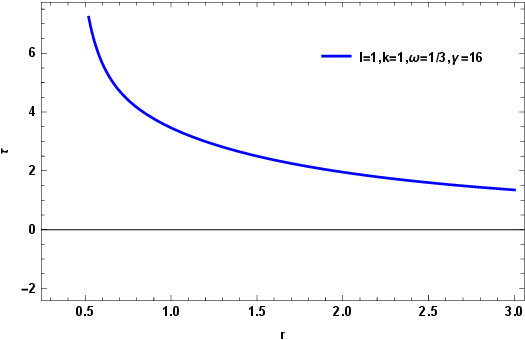}
 \label{2a}}
 \subfigure[]{
 \includegraphics[height=4.2cm,width=4.2cm]{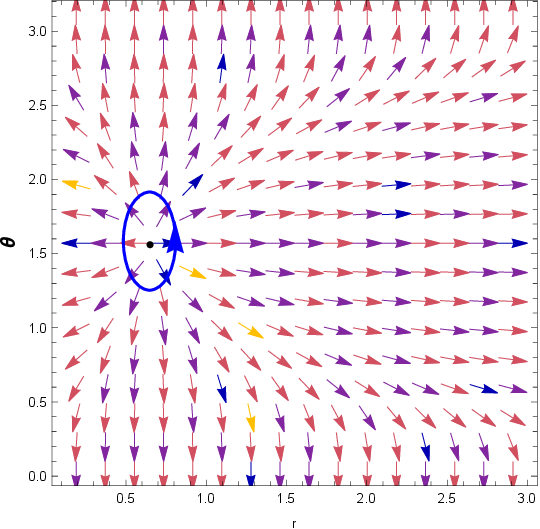}
 \label{2b}}
 \subfigure[]{
 \includegraphics[height=4.2cm,width=4.2cm]{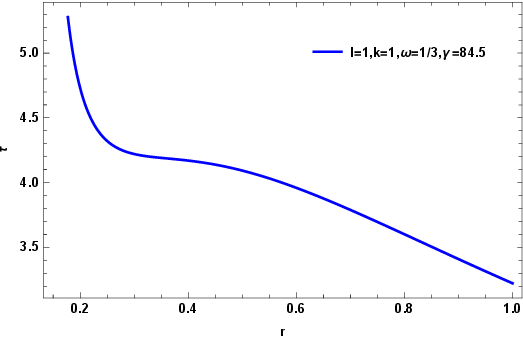}
 \label{2c}}
\subfigure[]{
 \includegraphics[height=4.2cm,width=4.2cm]{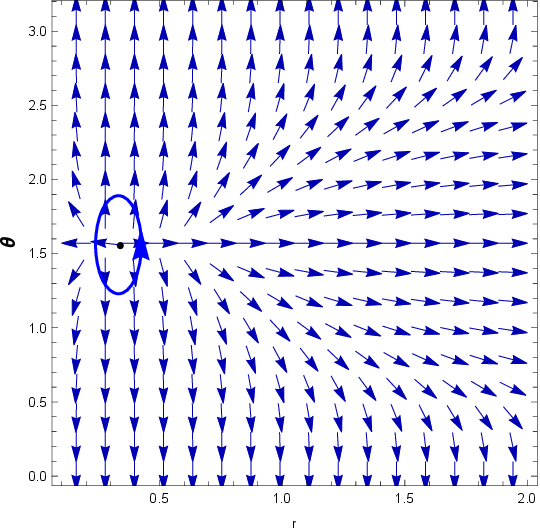}
 \label{2d}}
  \caption{\small{The curve corresponding to Eq.(\ref{16}) is depicted in Figs. \ref{2a} and \ref{2c}. In Figs. \ref{2b} and \ref{2d}, the ZPs are located at $(r, \theta)$ on the circular loops with parameters $(\ell=k=1, \omega=\frac{1}{3}, \gamma=16)$ and $(\ell=k=1, \omega=\frac{1}{3}, \gamma=-84.5)$. }}
 \label{F2}
 \end{center}
 \end{figure}

\begin{figure}[h!]
 \begin{center}
 \subfigure[]{
 \includegraphics[height=4.2cm,width=4.2cm]{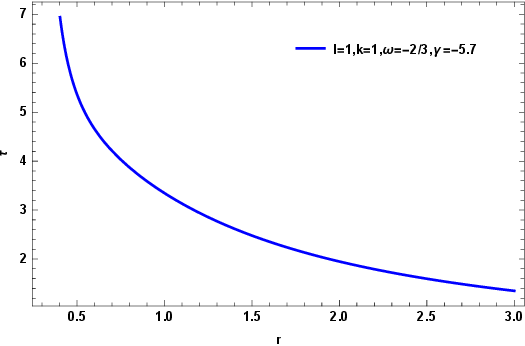}
 \label{3a}}
 \subfigure[]{
 \includegraphics[height=4.2cm,width=4.2cm]{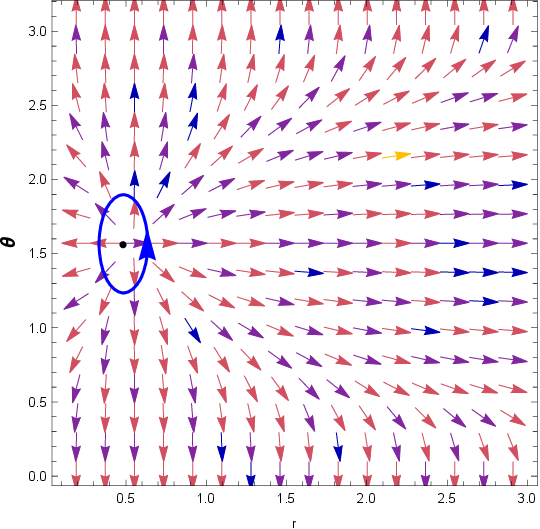}
 \label{3b}}
 \subfigure[]{
 \includegraphics[height=4.2cm,width=4.2cm]{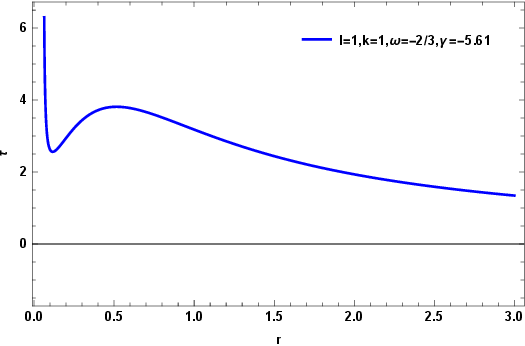}
 \label{3c}}
\subfigure[]{
 \includegraphics[height=4.2cm,width=4.2cm]{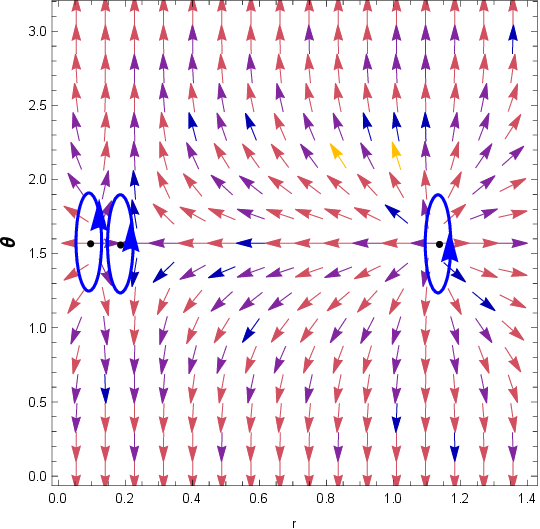}
 \label{3d}}
  \caption{\small{The curve corresponding to Eq.(\ref{16}) is depicted in Figs. \ref{3a} and \ref{3c}. In Figs. \ref{3b} and \ref{3d}, the ZPs are located at $(r, \theta)$ on the circular loops with parameters $(\ell=k=1, \omega=-\frac{2}{3}, \gamma=-5.7)$ and $(\ell=k=1, \omega=-\frac{2}{3}, \gamma=-5.61)$.}}
 \label{F3}
 \end{center}
 \end{figure}

\begin{figure}[h!]
 \begin{center}
 \subfigure[]{
 \includegraphics[height=4.2cm,width=4.2cm]{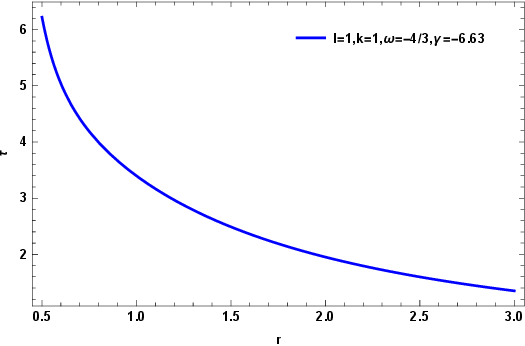}
 \label{4a}}
 \subfigure[]{
 \includegraphics[height=4.2cm,width=4.2cm]{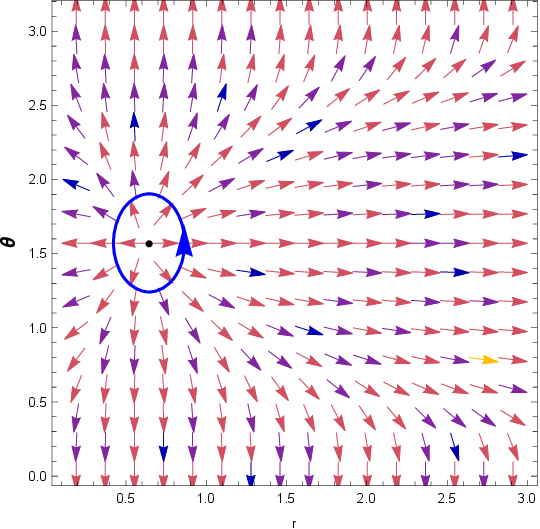}
 \label{4b}}
 \subfigure[]{
 \includegraphics[height=4.2cm,width=4.2cm]{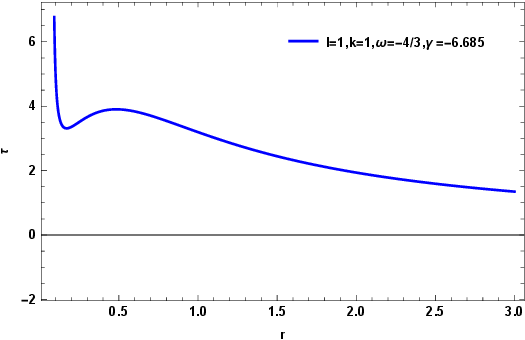}
 \label{4c}}
\subfigure[]{
 \includegraphics[height=4.2cm,width=4.2cm]{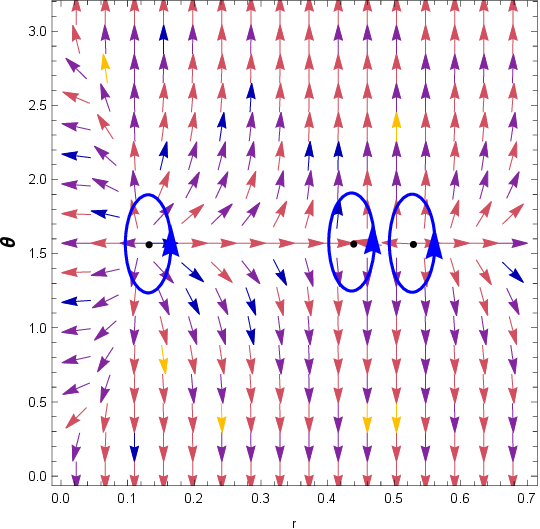}
 \label{4d}}
  \caption{\small{The curve corresponding to Eq.(\ref{16}) is depicted in Figs. \ref{4a} and \ref{4c}. In Figs. \ref{4b} and \ref{4d}, the ZPs are located at $(r, \theta)$ on the circular loops with parameters $(\ell=k=1, \omega=-\frac{4}{3}, \gamma=-6.63)$ and $(\ell=k=1, \omega=-\frac{4}{3}, \gamma=-6.685)$.}}
 \label{F4}
 \end{center}
 \end{figure}

To investigate the thermodynamic properties of black holes, various quantities are utilized. For instance, mass and temperature can describe the generalized free energy. Considering the relationship between mass and energy in black holes, we can represent our generalized free energy function as a standard thermodynamic function in the following form \cite{19}.
\begin{equation}\label{70}
\mathcal{F}=M-\frac{S}{\tau}.
\end{equation}
where $\tau$ denotes the Euclidean time period, while \( T \) (the inverse of $\tau$) represents the temperature of the ensemble. The generalized free energy is on-shell only when $\tau = \tau_{H} = \frac{1}{T_{H}}$. As stated in \cite{19}, a vector $\phi$ is constructed as follows,
\begin{equation}\label{8}
\phi=\big(\frac{\partial\mathcal{F}}{\partial r_{H}},-\cot\Theta\csc\Theta\big).
\end{equation}
Where $\phi^{\Theta}$ diverges, the vector direction points outward at $\Theta = 0$ and $\Theta = \pi$. The ranges for $r_{H}$ and $\Theta$ are $0 \leq r_{H} \leq \infty$ and $0 \leq \Theta \leq \pi$, respectively. Using Duan's $\phi$-mapping topological current theory, a topological current can be defined as follows,
\begin{equation}\label{9}
j^{\mu}=\frac{1}{2\pi}\varepsilon^{\mu\nu\rho}\varepsilon_{ab}\partial_{\nu}n^{a}\partial_{\rho}n^{b},\hspace{1cm}\mu,\nu,\rho=0,1,2
\end{equation}
Given \( n = (n^1, n^2) \), where \( n^1 = \frac{\phi^r}{\|\phi\|} \) and \( n^2 = \frac{\phi^\Theta}{\|\phi\|} \), Noether's theorem ensures that the resulting topological currents are conserved.
\begin{equation}\label{100}
\partial_{\mu}j^{\mu}=0,
\end{equation}
To determine the topological number, we reformulate the topological current \cite{19},
\begin{equation}\label{11}
j^{\mu}=\delta^{2}(\phi) J^{\mu}(\frac{\phi}{x}),
\end{equation}
The Jacobi tensor determine as,
\begin{equation}\label{12}
\varepsilon^{ab}J^{\mu}(\frac{\phi}{x})=\varepsilon^{\mu\nu\rho}\partial_{\nu}\phi^{a}\partial_{\rho}\phi^{b}.
\end{equation}
The Jacobi vector reduces to the standard Jacobi when $\mu=0$, as demonstrated by $J^{0}\left(\frac{\phi}{x}\right)=\frac{\partial(\phi^1,\phi^2)}{\partial(x^1,x^2)}$. Eq.(\ref{100}) shows that $j^{\mu}$ is only non-zero when $\phi=0$. Through some calculations, we can express the topological number or total charge $W$ as follows,
\begin{equation}\label{13}
W=\int_{\Sigma}j^{0}d^2 x=\Sigma_{i=1}^{n}\beta_{i}\eta_{i}=\Sigma_{i=1}^{n}\widetilde{\omega}_{i}.
\end{equation}
Here, $\beta_i$ denotes the positive Hopf index, which counts the loops of the vector $\phi^a$ in the $\phi$ space when $x^\mu$ is near the zero point $z_i$. Meanwhile, $\eta_i=\text{sign}(j^0(\phi/x)_{z_i})=\pm 1$. The quantity $\widetilde{\omega}_i$ represents the winding number for the $i$-th zero point of $\phi$ in $\Sigma$. Note that the winding number is independent of the shape of the region where the calculation occurs. The value of the winding number is directly related to black hole stability, with a positive (negative) winding number corresponding to a stable (unstable) black hole state. Using Eqs. (\ref{4}), (\ref{6}), and (\ref{70}), we derive the generalized Helmholtz free energy for Kiselev-AdS black holes within the framework of \( f(R, T) \) gravity.
\begin{equation}\label{14}
\mathcal{F}=\frac{1}{2} r \left(k r^{-\frac{8 \gamma  \omega +24 \pi  \omega +8 \pi }{\gamma  (-\omega )+3 \gamma +8 \pi }}+\frac{r^2}{l^2}+1\right)-\frac{\pi  r^2}{\tau }
\end{equation}
Based on the discussion in the previous section, the form of the function $(\phi^{r}, \phi^\Theta)$ is determined as follows,
\begin{equation}\label{15}
\begin{split}
&\phi ^{r}=\frac{3 k (-3 \gamma  \omega +\gamma -8 \pi  \omega ) r^{\frac{8 (\gamma  \omega +3 \pi  \omega +\pi )}{\gamma  (\omega -3)-8 \pi }}}{2 (8 \pi -\gamma  (\omega -3))}+\frac{3 r^2}{2 l^2}-\frac{2 \pi  r}{\tau }+\frac{1}{2}\\
&\phi ^{\theta }=-\frac{\cot (\theta )}{\sin (\theta )}
\end{split}
\end{equation}
The unit vectors \( \mathbf{n}_1 \) and \( \mathbf{n}_2 \) are computed using Eq.(\ref{15}). Next, we find the zero points of the \( \phi^{r} \) component by solving \( \phi^{r} = 0 \) and derive an expression for \( \tau \) as follows,
\begin{equation}\label{16}
\begin{split}
\tau =4 \pi  r\bigg[\frac{3 k (-3 \gamma  \omega +\gamma -8 \pi  \omega ) r^{\frac{8 (\gamma  \omega +3 \pi  \omega +\pi )}{\gamma  (\omega -3)-8 \pi }}}{8 \pi -\gamma  (\omega -3)}+\frac{3 r^2}{l^2}+1\bigg]^{-1}
\end{split}
\end{equation}
We face a single zero point in Figs. \ref{1b}, \ref{2b}, \ref{2d}, \ref{3b}, \ref{4b}, indicating one topological charge determined by the free parameters mentioned in the study. This charge corresponds to the winding number and is located within the blue contour loops at coordinates \((r, \theta)\). The sequence of the illustrations is determined by the parameters \(\omega\) and \(\gamma\); for example, in Fig. (\ref{F1}), we have (\(\omega = 0\) with \(\gamma = -1.9, -0.4\)), and for Figs. (\ref{F2}) through (\ref{F4}), \(\omega\) takes the values (\(\omega\)=\(\frac{1}{3}\) with \(\gamma = 16, 84.5\)), (\(\omega\)=\(-\frac{2}{3}\) with \(\gamma = -5.7, -5.61\)), and (\(\omega\)=\(-\frac{4}{3}\) with \(\gamma = -6.63, -6.685\)), respectively. In all Figs, we set \(k = \ell = 1\). The findings from all Figs show that the distinctive feature of total topological charge $+1$ is the zero point enclosed within the contour.\\

Additionally, our findings indicate that by increasing the parameter \(\gamma\) for \((\omega = 0, -\frac{2}{3})\), the number of topological charges increases, as shown in Figs. \ref{1b}, \ref{1d}, \ref{3b}, \ref{3d}, but the total topological charge remains \(W = +1\). Conversely, for \((\omega=-\frac{4}{3})\), the number of topological charges increases by decreasing the value of the parameter \(\gamma\). Generally, for \((\frac{1}{3})\), changing the parameter \(\gamma\) does not alter the number of topological charges.
However the the parameters \(\gamma\) and \(\omega\) affect the number of topological charges in the entire system, while the total topological charges for all modes are equal \(W = +1\).
Also,  Figs. \ref{1d}, \ref{3d}, \ref{4d} depict three topological charges \((\widetilde{\omega} = +1, -1, +1)\), resulting in a total topological charge of \(W = +1\). In Figs. \ref{1a}, \ref{1c}, \ref{2a}, \ref{2c}, \ref{3a}, \ref{3c}, \ref{4a}, \ref{4c}, we plotted the trajectory corresponding to Eq.(\ref{16}) across various free parameter values. The results are summarized in Table (\ref{1'}).

 \begin{center}
\begin{table}
  \centering
\begin{tabular}{|p{6cm}||p{3cm}||p{3cm}|}
  \hline
   \hspace{1.3cm}Free parameters  & \hspace{1.2cm} $\widetilde{\omega}$  & \hspace{1.2cm} $W$ \\[3mm]
   \hline
    $(\ell=k=1, \omega=0, \gamma=-1.9)$ &\hspace{1cm} $+1$ & \hspace{1cm} $+1$ \\[3mm]
   \hline
   $(\ell=k=1, \omega=0, \gamma=-0.4)$ & \hspace{0.7cm} $+1,-1,+1$ & \hspace{1cm}  $+1$ \\[3mm]
  \hline
   $(\ell=k=1, \omega=\frac{1}{3}, \gamma=16)$ & \hspace{1cm}  $+1$ & \hspace{1cm}  $+1$ \\[3mm]
  \hline
  $(\ell=k=1, \omega=\frac{1}{3}, \gamma=84.5)$& \hspace{1cm} $+1$ & \hspace{1cm} $+1$  \\[3mm]
  \hline
  $(\ell=k=1, \omega=-\frac{2}{3}, \gamma=-5.7)$ & \hspace{1cm} $+1$ & \hspace{1cm} $+1$ \\[3mm]
  \hline
  $(\ell=k=1, \omega=-\frac{2}{3}, \gamma=-5.61)$ & \hspace{0.8cm} $+1,-1,+1$ & \hspace{1cm} $+1$ \\[3mm]
  \hline
  $(\ell=k=1, \omega=-\frac{4}{3}, \gamma=-6.63)$ & \hspace{1cm} $+1$ & \hspace{1cm} $+1$ \\[3mm]
  \hline
  $(\ell=k=1, \omega=-\frac{4}{3}, \gamma=-6.685)$ & \hspace{0.8cm} $+1,-1,+1$ & \hspace{1cm} $+1$ \\
  \hline
\end{tabular}
\caption{Summary of the results for F-Model}\label{1'}
\end{table}
 \end{center}
\newpage
\subsubsection{T-Method}
\begin{figure}[h!]
 \begin{center}
 \subfigure[]{
 \includegraphics[height=4cm,width=4cm]{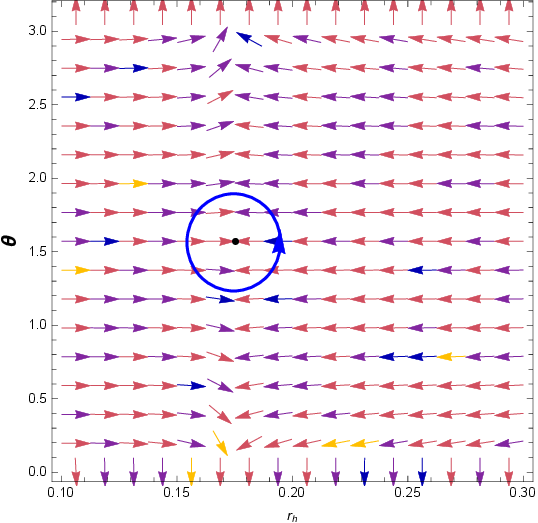}
 \label{5a}}
 \subfigure[]{
 \includegraphics[height=4cm,width=4cm]{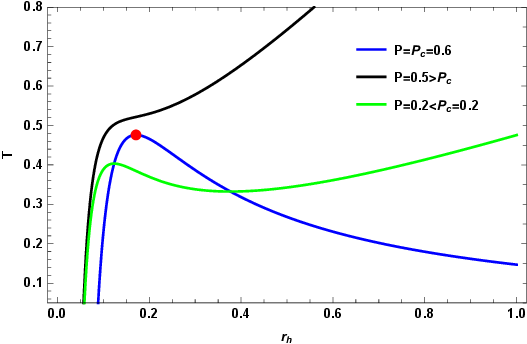}
 \label{5b}}
 \subfigure[]{
 \includegraphics[height=4cm,width=4cm]{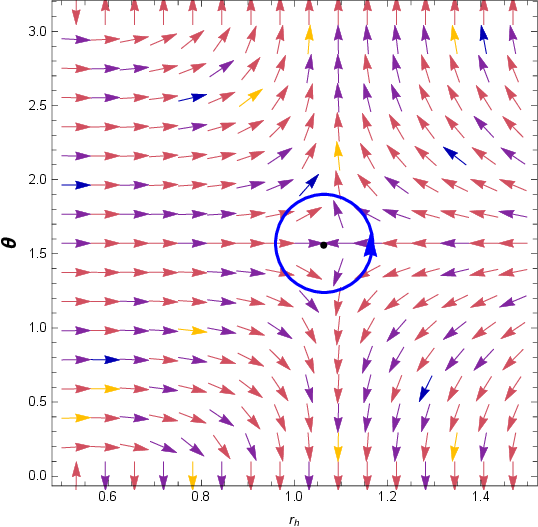}
 \label{5c}}
 \subfigure[]{
 \includegraphics[height=4cm,width=4cm]{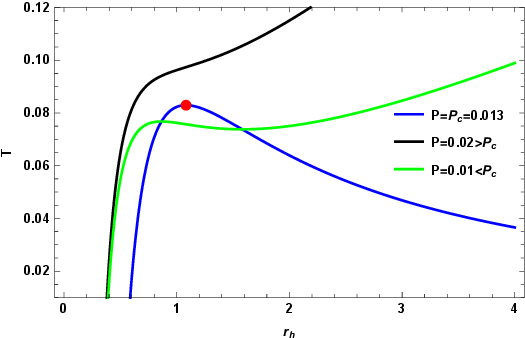}
 \label{5d}}
  \caption{\small{The arrows represent the vector field n on the $r_H-\theta$ plane for the Kiselev-AdS black holes within f (R, T)  gravity with $(k =\ell= 1, \omega = 0, \gamma = -0.4)$ and $(k =\ell= 1, \omega = 0, \gamma = -1.9)$ in Figs. \ref{5a} and \ref{5c}, respectively.
Isobaric and spinodal curves (blue lines) for the Kiselev-AdS Black Holes within f (R, T)  gravity in Figs. \ref{5b} and \ref{5d}}}
 \label{T1}
 \end{center}
 \end{figure}

\begin{figure}[h!]
 \begin{center}
 \subfigure[]{
 \includegraphics[height=4cm,width=4cm]{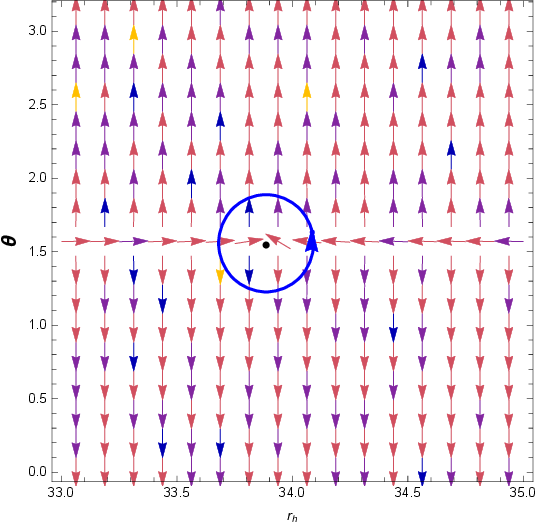}
 \label{6a}}
 \subfigure[]{
 \includegraphics[height=4cm,width=4cm]{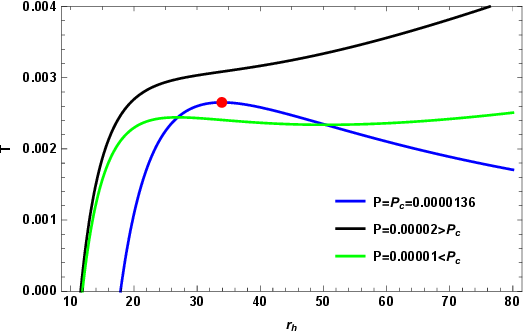}
 \label{6b}}
 \subfigure[]{
 \includegraphics[height=4cm,width=4cm]{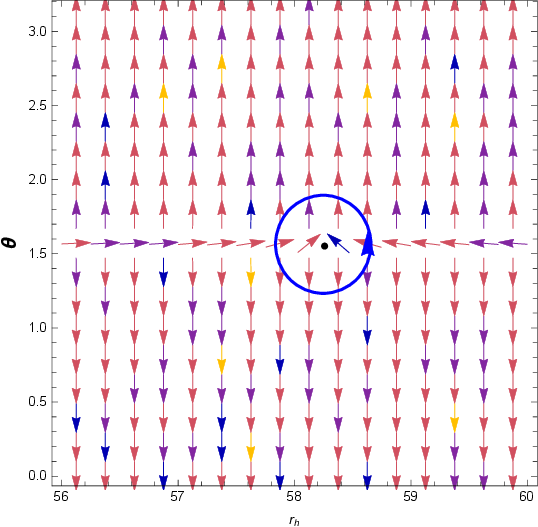}
 \label{6c}}
 \subfigure[]{
 \includegraphics[height=4cm,width=4cm]{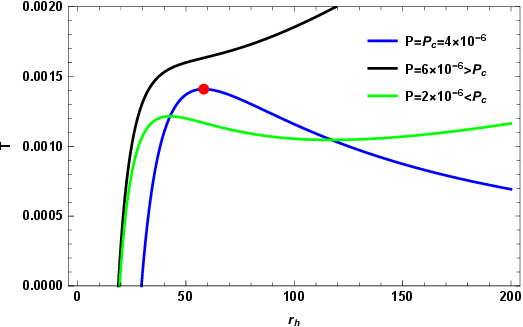}
 \label{6d}}
  \caption{\small{The arrows represent the vector field n on the $r_H-\theta$ plane for the mentioned black hole with $(k =\ell= 1, \omega = -4/3, \gamma = -6.63)$ and $(k =\ell= 1, \omega = -4/3, \gamma = -6.685)$ in Figs. \ref{6a} and \ref{6c}, respectively.
Isobaric and spinodal curves (blue lines) in Figs. \ref{6b} and \ref{6d}}}
 \label{T2}
 \end{center}
 \end{figure}

In the extended thermodynamic framework, the temperature \(T\) of a system like a black hole is expressed as a function of pressure \(P\), and an additional parameter \(\xi\). Critical points are identified using the conditions,
\begin{equation}\label{17}
\begin{split}
\left(\frac{\partial T}{\partial r_{H}}\right)_{P, \xi} = 0, \quad \left(\frac{\partial^2 T}{\partial r_{H}^2}\right)_{P, \xi} = 0.
\end{split}
\end{equation}
Recent research indicates that each critical point can have a topological charge, which can be conventional or novel. To eliminate the parameter \(P\), the relation:$
\left(\frac{\partial T}{\partial r_{H}}\right)_{P, \xi} = 0
$ is used. To study the topological charge, the thermodynamic function \(\Phi\) is defined as,
\begin{equation}\label{18}
\begin{split}
\Phi = \frac{1}{\sin \theta} T(r_H, \xi),
\end{split}
\end{equation}
where \(\frac{1}{\sin \theta}\) simplifies the calculations. A new vector field \(\phi = (\phi^{r_H}, \phi^\theta)\) is defined using Duan’s \(\phi\)-mapping theory,
\begin{equation}\label{19}
\begin{split}
\phi^{rH} = \left(\frac{\partial \Phi}{\partial r_{H}}\right)_{\theta, \xi}, \quad \phi^\theta = \left(\frac{\partial \Phi}{\partial \theta}\right)_{r, \xi}.
\end{split}
\end{equation}
The vector field \(\phi\) is zero at \(\theta = \frac{\pi}{2}\), helping to identify critical points. The points \(\theta = 0\) and \(\theta = \pi\) serve as boundaries in the parameter space. When the vector field \(\phi^a\) is zero, its topological current \(j^\mu\) becomes non-zero. Conversely, contour \( C \) can be parameterized by \( \vartheta \in (0, 2\pi) \) with \( r = a \cos \vartheta + r_0 \) and \( \theta = b \sin \vartheta + \frac{\pi}{2} \), where \( (r_0, \frac{\pi}{2}) \) represents the center of the contour. We introduce a new quantity to measure the deflection of the vector field along the contour,
\begin{equation}\label{20}
\begin{split}
\Omega(\vartheta) = \int_0^{\vartheta} \epsilon_{ij} n^i \frac{\partial n^j}{\partial \vartheta} d\vartheta,
\end{split}
\end{equation}
where \( i, j = S, \theta \). Consequently, the topological charge \( Q_t \) is given by,
\begin{equation}\label{21}
\begin{split}
Q_t = \frac{\Omega(2\pi)}{2\pi}.
\end{split}
\end{equation}
The temperature is reformulated by eliminating the pressure parameter,
\begin{equation}\label{22}
\begin{split}
&T=\frac{1}{4 \pi }\bigg[\frac{3 k (\gamma -(3+8 \pi ) \omega ) (\gamma  (7 \omega +3)+8 \pi  (3 \omega +2)) r_{H}^{1-\frac{6 (\gamma +4 \pi ) (\omega +1)}{8 \pi -\gamma  (\omega -3)}}}{(\gamma  (\omega -3)-8 \pi )^2}\\
&+\frac{3 k (\gamma -(3+8 \pi ) \omega ) r_{H}^{\frac{\gamma  (7 \omega +3)+8 \pi  (3 \omega +2)}{\gamma  (\omega -3)-8 \pi }}}{8 \pi -\gamma  (\omega -3)}+\frac{2}{r_{H}}\bigg]
\end{split}
\end{equation}
Subsequently, by substituting equation (\ref{22}) into equation (\ref{18}), the thermodynamic function $\Phi $ for the black hole can be derived as follows,
\begin{equation}\label{23}
\begin{split}
&\Phi =\frac{1}{4 \pi \sin\theta}\bigg[\frac{3 k (\gamma -(3+8 \pi ) \omega ) (\gamma  (7 \omega +3)+8 \pi  (3 \omega +2)) r_{H}^{1-\frac{6 (\gamma +4 \pi ) (\omega +1)}{8 \pi -\gamma  (\omega -3)}}}{(\gamma  (\omega -3)-8 \pi )^2}\\
&+\frac{3 k (\gamma -(3+8 \pi ) \omega ) r_{H}^{\frac{\gamma  (7 \omega +3)+8 \pi  (3 \omega +2)}{\gamma  (\omega -3)-8 \pi }}}{8 \pi -\gamma  (\omega -3)}+\frac{2}{r_{H}}\bigg]
\end{split}
\end{equation}
Topological charges can be identified due to the presence of critical points within acceptable regions for free parameters. By applying Eqs. (\ref{17}), (\ref{18}), (\ref{19}), (\ref{22}), and (\ref{23}) for \(\omega = 0\) and \(\omega = -4/3\) with various values of \(\gamma\) as shown in Figs. \ref{5a}, \ref{5c}, \ref{6a} and \ref{6c}, the topological charges of each critical point can be determined. Consequently, from Figs. (\ref{T1}) and (\ref{T2}), we find that \(Q_{CP1} = -1\). Additionally, we conclude that \(Q_{total} = -1\).

As illustrated in Figs. \ref{5b}, \ref{5d}, \ref{6b} and \ref{6d}, the conventional critical points represent the local minima and maxima of the spinodal curve. The range of various parameters for the black hole in question is constrained by the conditions of the event horizon, which has been thoroughly examined in \cite{48}. Therefore, for other values of the parameter \(\omega\), specifically \(1/3\) and \(-2/3\), the topological charges were not obtained within this range. Consequently, the discussed topics apply only to (dust) and (phantom) fields.

\subsection{Photon spheres of Kiselev-AdS black holes in \( f(R, T) \) gravity}
Referring to \cite{55,56,57,58}, we start with a regular potential,
\begin{equation}\label{Ph1}
\begin{split}
H(r,\theta)=\sqrt{\frac{-g_{tt}}{g_{\varphi\varphi}}}=\frac{1}{\sin\theta}\left(\frac{f(r)}{h(r)}\right)^{1/2},
\end{split}
\end{equation}
By examining this potential, we can determine the radius of the photon sphere, which is located at,
\begin{equation}\label{Ph2}
\begin{split}
\partial_r H=0,
\end{split}
\end{equation}
We then introduce a vector field \( \phi = (\phi^r, \phi^\theta) \), defined as follows,
\begin{equation}\label{Ph3}
\begin{split}
\phi^r=\frac{\partial_r H}{\sqrt{g_{rr}}}=\sqrt{g(r)}\partial_r H, \quad \phi^\theta=\frac{\partial_\theta H}{\sqrt{g_{\theta\theta}}}=\frac{\partial_\theta H}{\sqrt{h(r)}},
\end{split}
\end{equation}
Consequently, the total charge is given by,
\begin{equation}\label{Ph4}
\begin{split}
Q=\sum_{i}\widetilde{\omega}_i,
\end{split}
\end{equation}
In conclusion, the presence of a zero point within a closed curve indicates that the charge \( Q \) is precisely equal to the winding number. For more study please see \cite{57}. Now with respect to B(r) in Eq. \ref{200} the above functions will be as follows,
\begin{equation}\label{Ph5}
H =\frac{\sqrt{1-\frac{2 M}{r}+\frac{r^{2}}{l^{2}}+\frac{k}{r^{\frac{8 \left(3 \omega  \pi +\gamma  \omega +\pi \right)}{-\gamma  \omega +8 \pi +3 \gamma}}}}}{\sin \! \left(\theta \right) r},
\end{equation}
\begin{equation}\label{Ph6}
\begin{split}
&\mathcal{A}=\left(-\frac{1}{12} \gamma  \omega +\frac{2}{3} \pi +\frac{1}{4} \gamma \right) r^{\frac{24 \omega  \pi +7 \gamma  \omega +16 \pi +3 \gamma}{-\gamma  \omega +8 \pi +3 \gamma}}\\
&\mathcal{B} =-2 \left(-\frac{1}{8} \gamma  \omega +\pi +\frac{3}{8} \gamma \right) M \,r^{\frac{\left(24 \pi +8 \gamma \right) \omega +8 \pi}{\left(-\omega +3\right) \gamma +8 \pi}}\\
&\mathcal{C} =\left(\pi +\frac{\gamma}{4}\right) \left(\omega +1\right) r k \\
&\phi^{r}=-\frac{12 \left(A +B +C \right) r^{\frac{\left(-5 \omega -9\right) \gamma -24 \omega  \pi -32 \pi}{\left(-\omega +3\right) \gamma +8 \pi}} \csc \! \left(\theta \right)}{-\gamma  \omega +8 \pi +3 \gamma},
\end{split}
\end{equation}
\begin{equation}\label{Ph7}
\phi^{\theta}=-\frac{\sqrt{1-\frac{2 M}{r}+\frac{r^{2}}{l^{2}}+\frac{k}{r^{\frac{8 \left(3 \omega  \pi +\gamma  \omega +\pi \right)}{-\gamma  \omega +8 \pi +3 \gamma}}}}\, \cos \! \left(\theta \right)}{\sin \! \left(\theta \right)^{2} r^{2}}.
\end{equation}
Since in this article, our main goal is to investigate the effect of gravitational corrections on the model, for this aim, we will study the structure of the photon sphere and its parameter range in three parts. It should be noted that for the sake of simplicity, we consider the values of $M=1, \ell=1, k=1 $ for all cases in this section.
\subsubsection{Case I: Dust field}
\begin{figure}[H]
 \begin{center}
 \subfigure[]{
 \includegraphics[height=5cm,width=8cm]{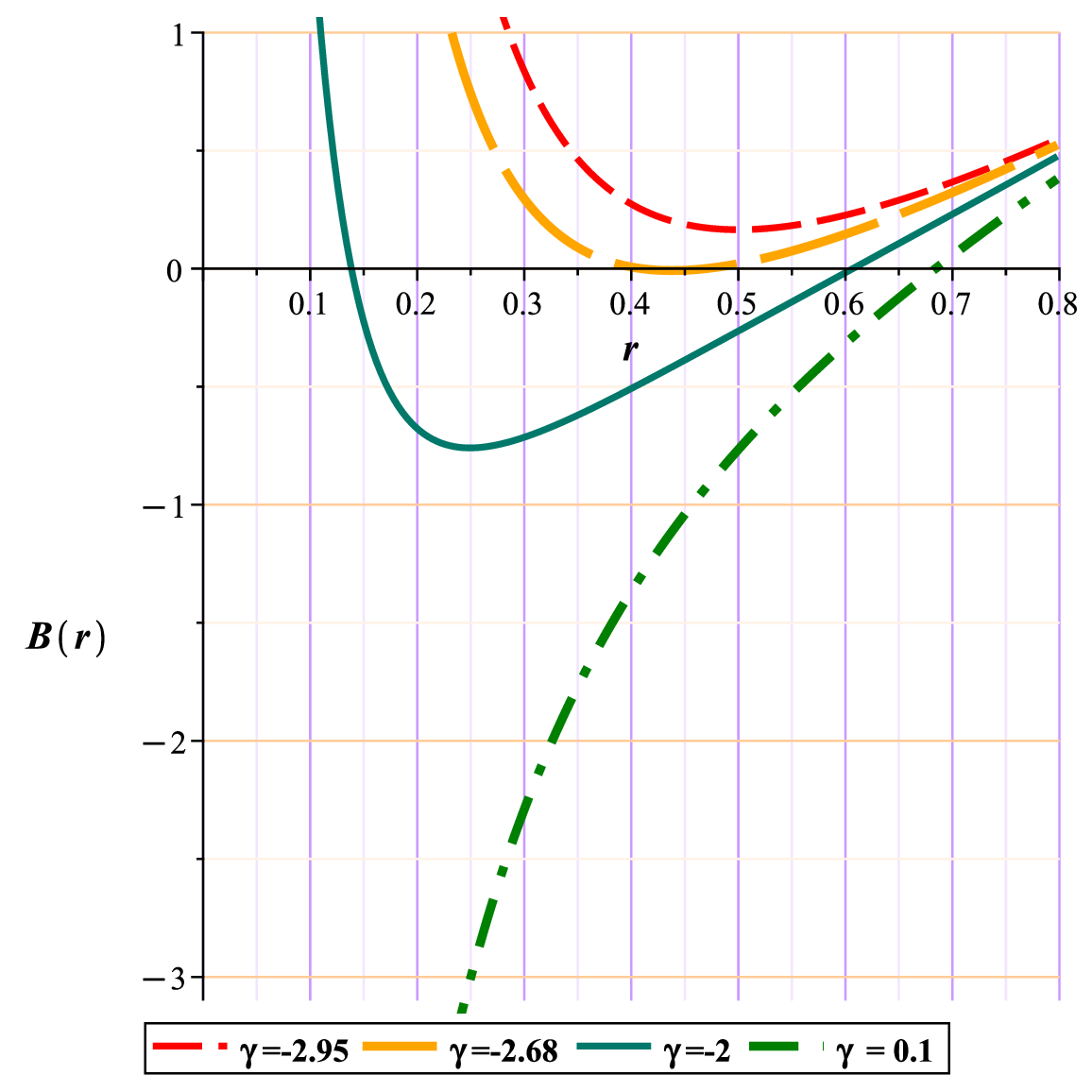}
 \label{7a}}
  \caption{{Metric function with different $\gamma$ for Kiselev-AdS black holes in \( f(R, T) \) gravity}}
 \label{7}
 \end{center}
 \end{figure}

We have drawn the metric function for $\omega=0$ and different $\gamma$. As can be seen in Fig. (\ref{7}), the metric function for $ \gamma>-2.6911 $ always has a root.
For this case, the general form of the  Eqs.(\ref{Ph1}) and (\ref{Ph3}) will be as follows,
 \begin{equation}\label{Ph8}
H =\frac{\sqrt{1-\frac{2}{r}+r^{2}+r^{-\frac{8 \pi}{8 \pi +3 \gamma}}}}{\sin \! \left(\theta \right) r},
\end{equation}
 \begin{equation}\label{Ph9}
\phi^{r}=-\frac{8 \csc \! \left(\theta \right) \left(\left(\frac{3 \pi}{2}+\frac{3 \gamma}{8}\right) r^{\frac{3 \gamma}{8 \pi +3 \gamma}}+\left(\pi +\frac{3 \gamma}{8}\right) \left(r -3\right)\right)}{r^{3} \left(8 \pi +3 \gamma \right)},
\end{equation}
\begin{equation}\label{Ph10}
\phi^{\theta}=-\frac{\sqrt{1-\frac{2}{r}+r^{2}+r^{-\frac{8 \pi}{8 \pi +3 \gamma}}}\, \cos \! \left(\theta \right)}{\sin \! \left(\theta \right)^{2} r^{2}}.
\end{equation}
 \begin{figure}[H]
 \begin{center}
 \subfigure[]{
 \includegraphics[height=5cm,width=8cm]{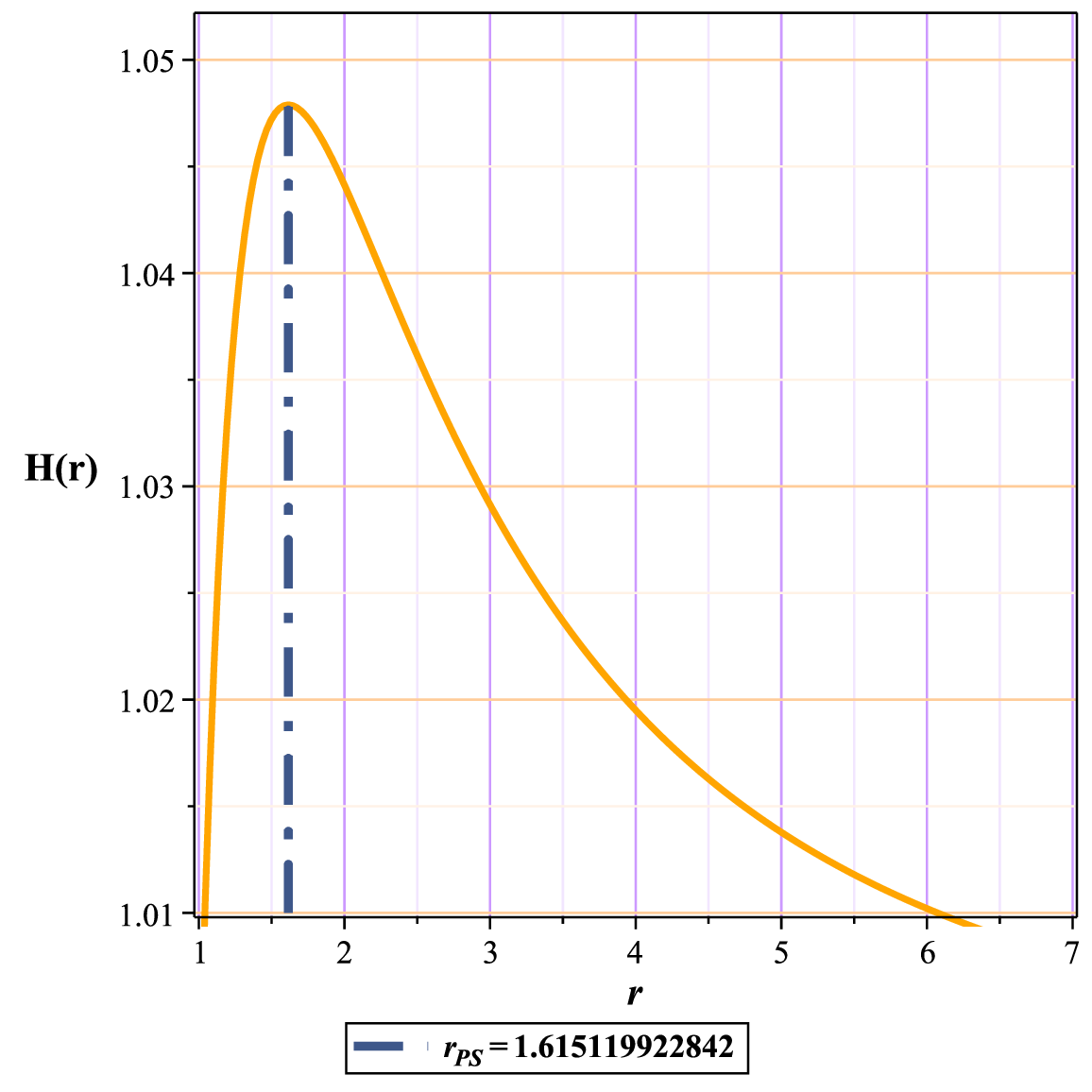}
 \label{8a}}
 \subfigure[]{
 \includegraphics[height=5cm,width=8cm]{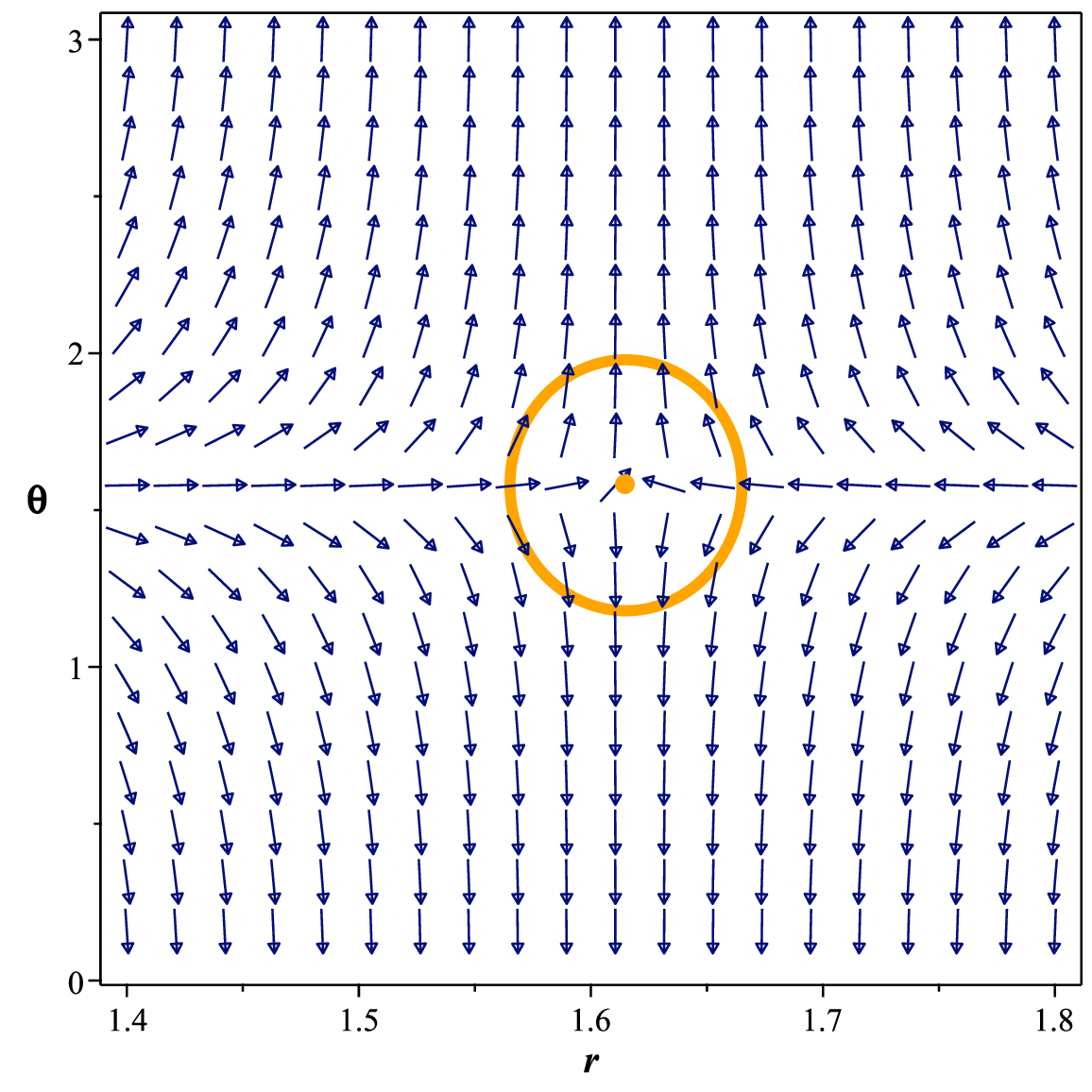}
 \label{8b}}
   \caption{\small{Fig. \ref{8a}: The topological potential H(r) for Kiselev-AdS black holes in f(R, T) gravity, \ref{8b}: The normal vector field $n$ in the $(r-\theta)$ plane. The photon sphere are located at $ (r,\theta)=(1.615119922842,1.57)$  with respect to $( \gamma=-2.68, m=1,l=1,k=1 )$ }}
 \label{8}
\end{center}
\end{figure}
 \begin{figure}[H]
 \begin{center}
 \subfigure[]{
 \includegraphics[height=5cm,width=8cm]{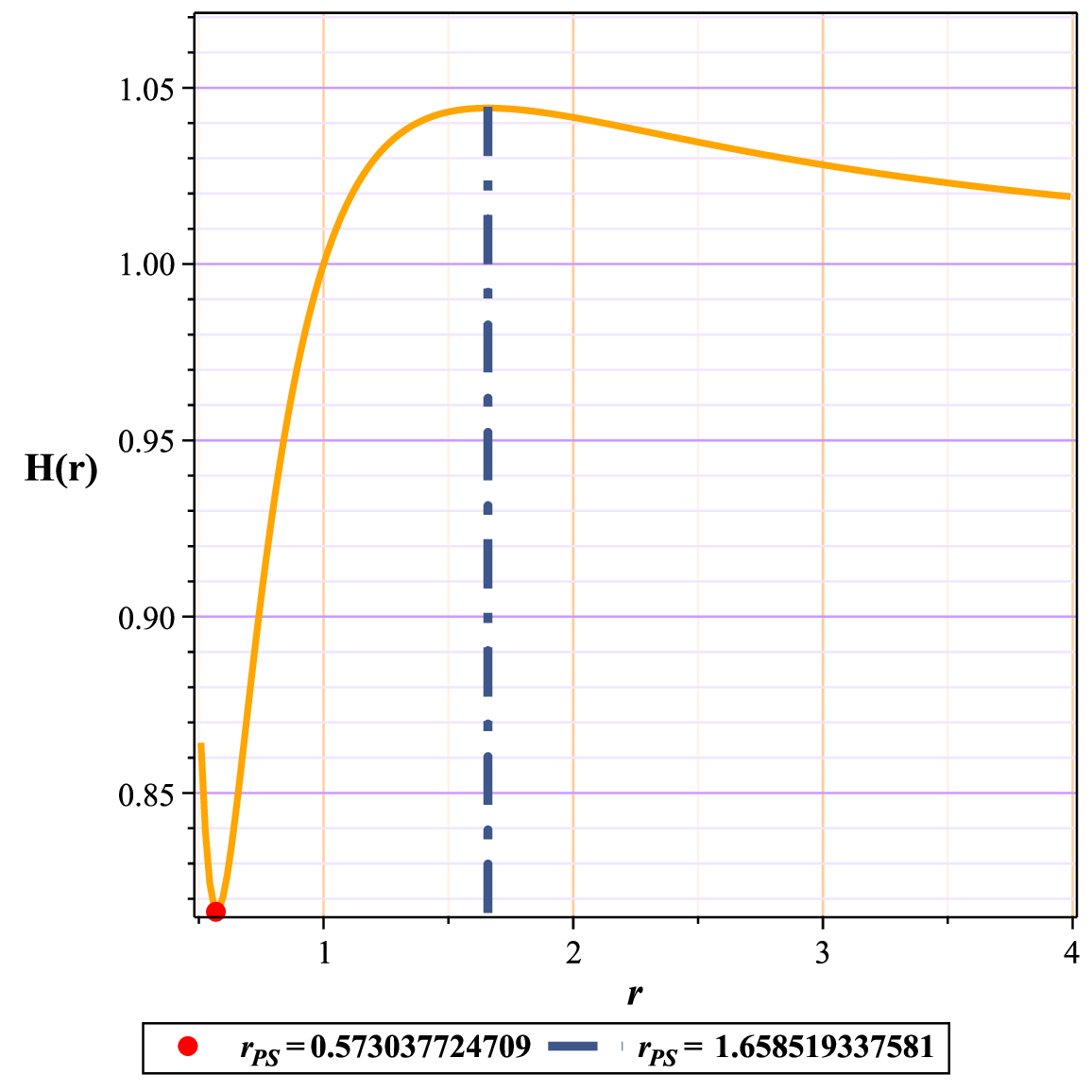}
 \label{9a}}
 \subfigure[]{
 \includegraphics[height=5cm,width=8cm]{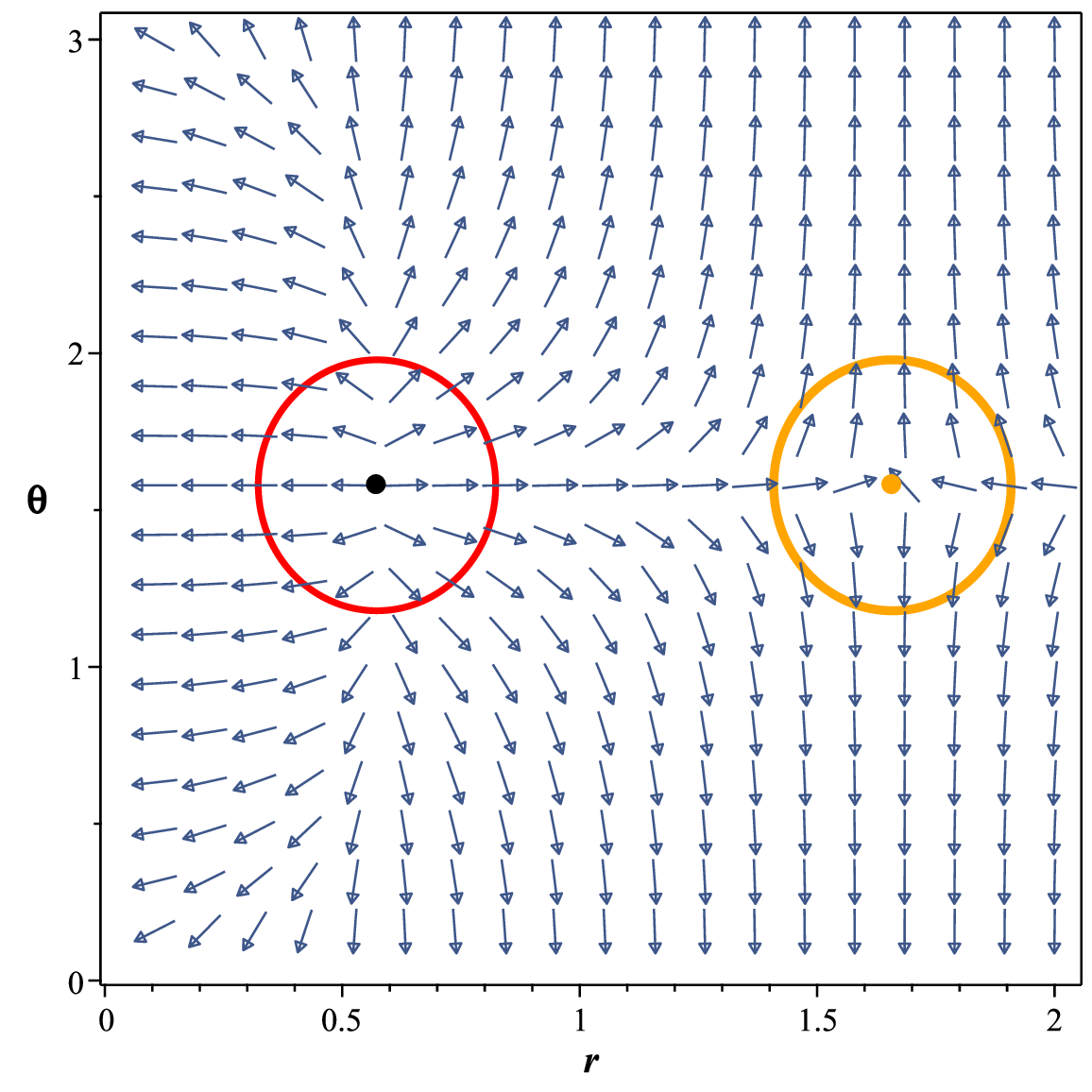}
 \label{9b}}
\caption{\small{Fig \ref{9a}: The topological potential H(r) for Kiselev-AdS Black Holes, \ref{9b}: The normal vector field $n$ in the $(r-\theta)$ plane. The photon sphere are located at $ (r,\theta)=(0.573037724709,1.57),(r,\theta)=(1.658519337581,1.57)$  with respect to $( \gamma=-3, m=1,l=1,k=1 )$ }}
 \label{9}
\end{center}
\end{figure}
\begin{center}
\begin{table}
  \centering
\begin{tabular}{|p{3cm}|p{4cm}|p{5cm}|p{1.5cm}|p{3cm}|}
  \hline
  \centering{Kiselev-AdS Black Holes}  & \centering{Fix parametes} &\centering{Conditions}& TTC&\ $(R_{PLPS})$\\[3mm]
   \hline
  \centering{unstable photon sphere} & \centering $k=1,m=1,l=1$ & \centering{$-2.6911\geq\gamma  $} &\centering $-1$&\ $1.616399193583$ \\[3mm]
   \hline
   \centering{naked singularity} & \centering $k=1,m=1,l=1$ & \centering{$-8.37<\gamma < -2.6911$} & \centering $ 0 $ &\ $-$ \\[3mm]
   \hline
   \end{tabular}
   \caption{$R_{PLPS}$: The minimum or maximum possible radius for the appearance of an unstable photon sphere. TTC: Total Topological Charge}\label{2'}
\end{table}
 \end{center}
Based on table (\ref{2'}), as evident from Figs. (\ref{7}), (\ref{8}) and (\ref{9}), in the first region, the potential structure possesses only a single local maximum or an unstable photon sphere with a negative unit charge. However, in the second region, with the disappearance of the event horizon, a local minimum (stable photon sphere) emerges beyond the horizon. This results in the total topological charge becoming zero, rendering the space in the form of a naked singularity, which is also confirmed by the metric function.\\
An interesting aspect of this model, in comparison to previous works, is the elimination of the forbidden region. In this model, beyond the singularity region, contrary to previous models where a forbidden zone existed, the metric function exhibits a single root at $\gamma = -8.38$, and a photon sphere appears. However, whether these $\gamma$ values are permissible within Einstein's equations and whether they satisfy the necessary energy conditions requires further scrutiny.
 \subsubsection{Case II: Quintessence field}
\begin{figure}[H]
 \begin{center}
 \subfigure[]{
 \includegraphics[height=5cm,width=5cm]{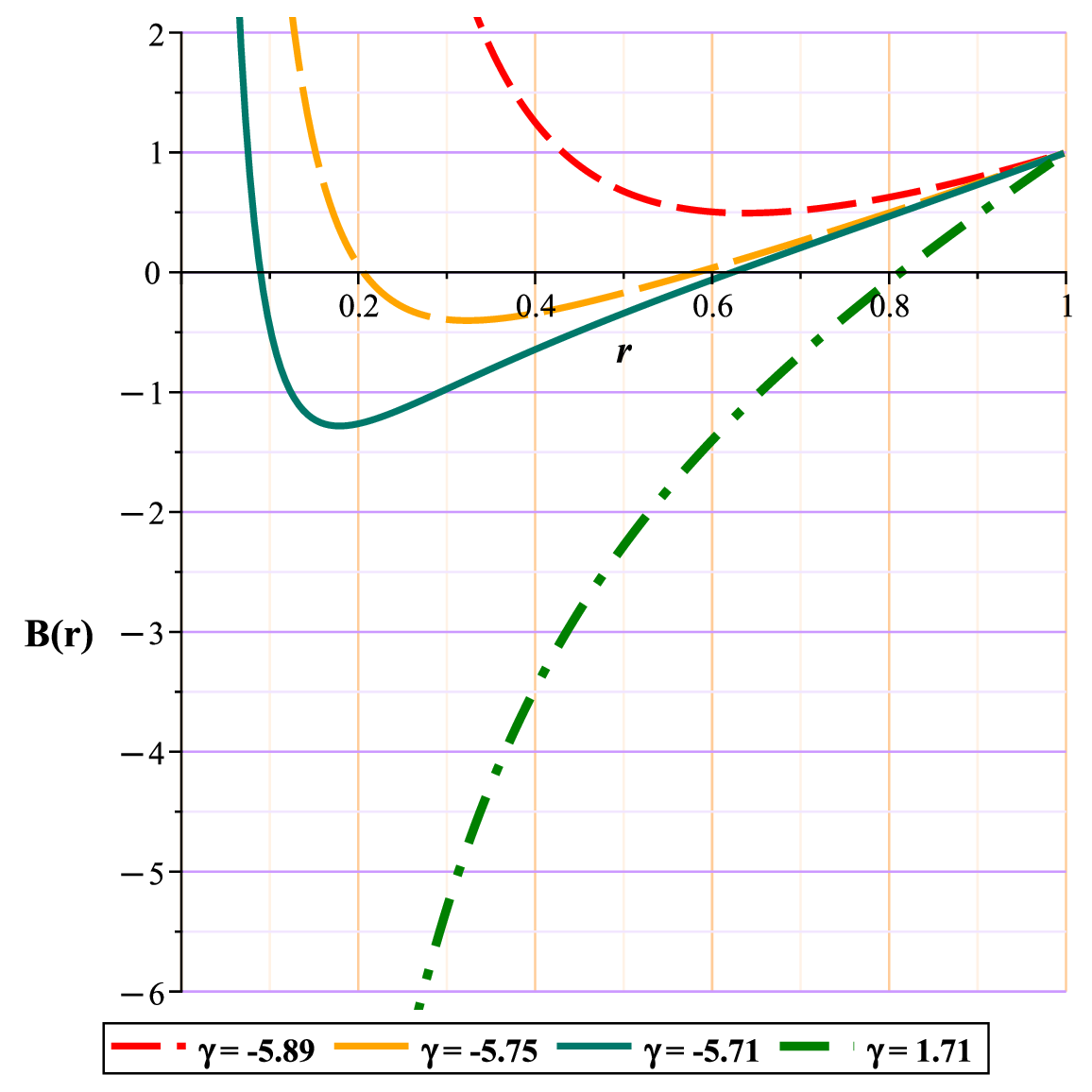}
 \label{11a}}
 \subfigure[]{
 \includegraphics[height=5cm,width=5cm]{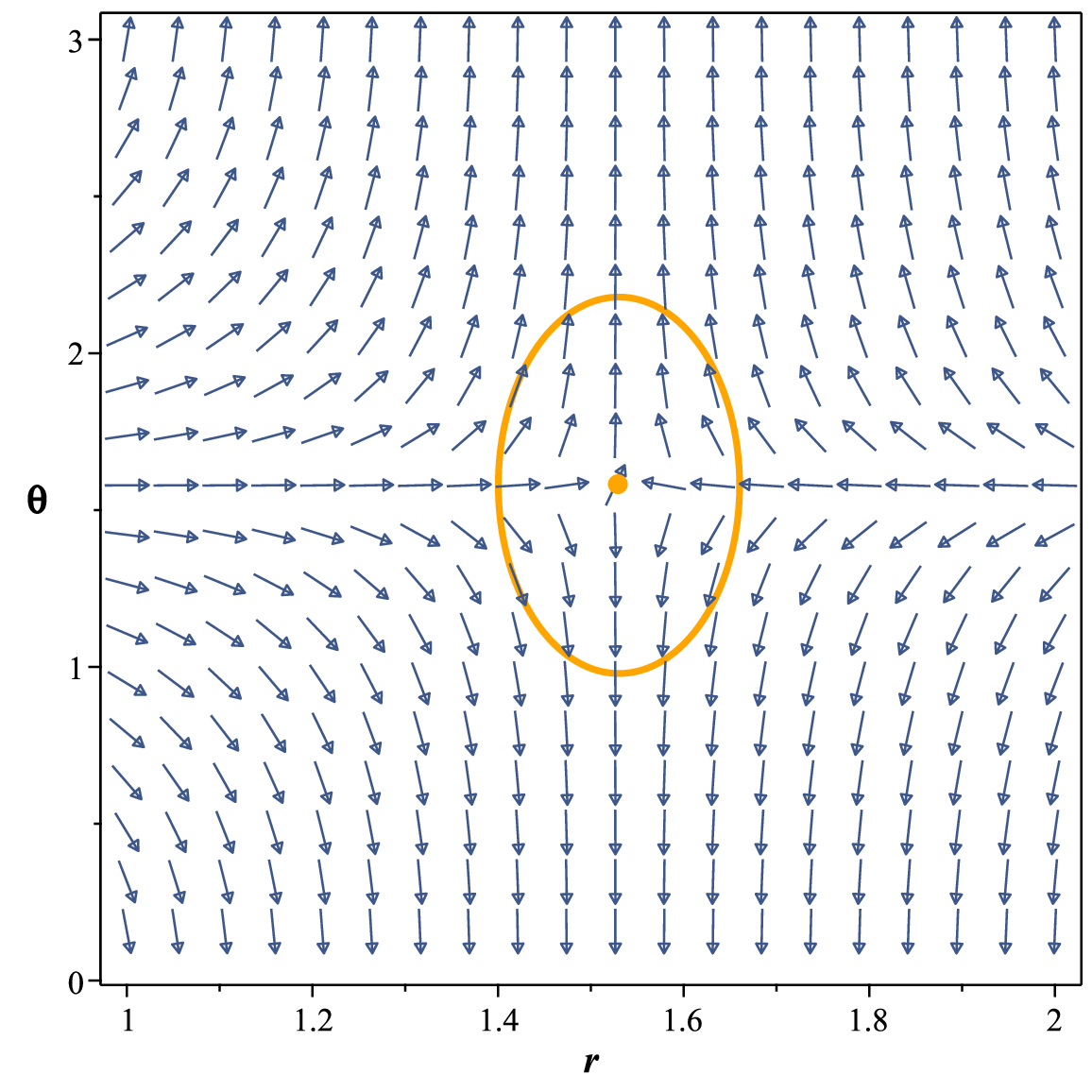}
 \label{11b}}
 \subfigure[]{
 \includegraphics[height=5cm,width=5cm]{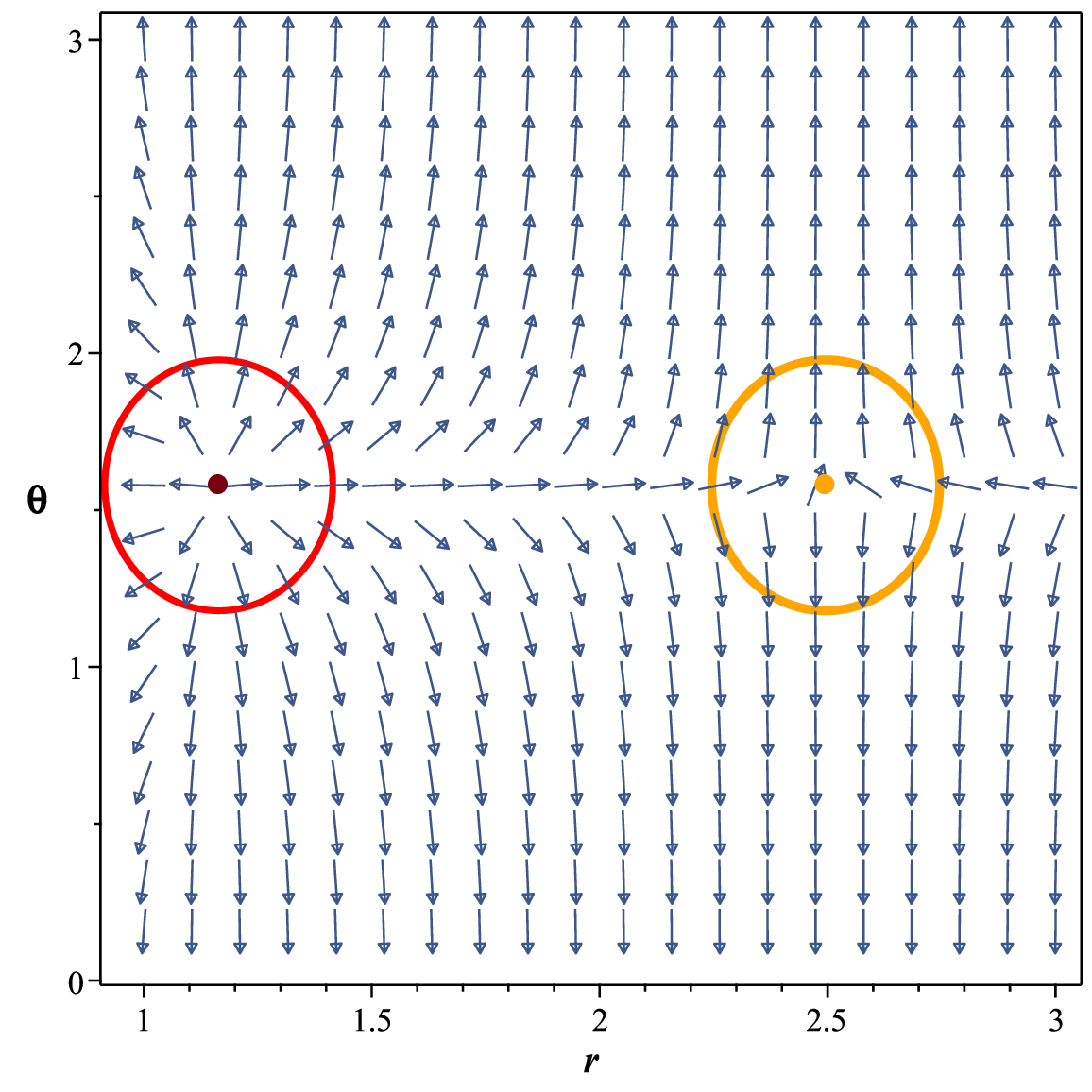}
 \label{11c}}
\caption{\small{Fig. \ref{11a}: Metric function with different $\gamma$ for Kiselev-AdS black holes in f(R, T) gravity, \ref{11b}: The normal vector field $n$ in the $(r-\theta)$ plane. The photon sphere are located at $ (r,\theta)=(1.53024,1.57)$  with respect to $( \gamma=-5.68, m=1,l=1,k=1 )$. \ref{11c}: The normal vector field $n$ in the $(r-\theta)$ plane. The photon sphere are located at $ (r,\theta)=(1.164619828617,1.57),(r,\theta)=(2.495766445262,1.57)$  with respect to $( \gamma=-6.1, m=1,l=1,k=1 )$}}
 \label{11}
\end{center}
\end{figure}

As can be seen in Fig. (\ref{11a}), with respect to $\omega=-2/3$ the metric function for $ \gamma>-5.7903 $ always has a root. Now with respect to Eqs. (\ref{Ph1}) and (\ref{Ph2}) we have,
\begin{equation}\label{Ph11}
H =\frac{\sqrt{\frac{k \,r^{\frac{24 \pi +16 \gamma}{24 \pi +11 \gamma}} r +r^{3}+r -2}{r}}}{\sin \! \left(\theta \right) r},
\end{equation}
 \begin{equation}\label{Ph12}
\phi^{r}=-\frac{12 \csc \! \left(\theta \right) \left(\left(\pi +\frac{\gamma}{4}\right) r^{\frac{48 \pi +27 \gamma }{24 \pi +11 \gamma}}+2 \left(r -3\right) \left(\pi +\frac{11 \gamma}{24}\right)\right)}{r^{3} \left(24 \pi +11 \gamma \right)},
\end{equation}
 \begin{equation}\label{Ph13}
\phi^{\theta}=-\frac{\sqrt{\frac{k \,r^{\frac{24 \pi +16 \gamma}{24 \pi +11 \gamma}} r +r^{3}+r -2}{r}}\, \cos \! \left(\theta \right)}{r^{2} \sin \! \left(\theta \right)^{2}}.
\end{equation}
The behavior of the model based on the ranges obtained for the $\gamma$ parameter is fully displayed in table (\ref{3'}) we have drawn a sample from each area in Fig. (\ref{11}).
\begin{center}
\begin{table}
  \centering
\begin{tabular}{|p{3cm}|p{4cm}|p{5cm}|p{1.5cm}|p{3cm}|}
  \hline
  \centering{Kiselev-AdS Black Holes}  & \centering{Fix parametes} &\centering{Conditions}& TTC&\ $(R_{PLPS})$\\[3mm]
   \hline
  \centering{unstable photon sphere} & \centering $k=1,m=1,l=1$ & \centering{$-5.7902\geq\gamma  $} &\centering $-1$&\ $1.616366040729$ \\[3mm]
   \hline
   \centering{naked singularity} & \centering $k=1,m=1,l=1$ & \centering{$-6.8544<\gamma < -5.7902$} & \centering $ 0 $ &\ $-$ \\[3mm]
   \hline
   \end{tabular}
   \caption{$R_{PLPS}$: The minimum or maximum possible radius for the appearance of an unstable photon sphere. TTC: Total Topological Charge}\label{3'}
\end{table}
 \end{center}
 \subsubsection{Case III: Phantom field}
In this scenario, the impact of corrections operates in the exact opposite manner compared to previous cases. Unlike the previous instances where corrections influenced the system by increasing the $\gamma$, in this case, it is the reduction of $\gamma$ that has a more significant effect. With respect to $\omega=-4/3$ the metric function for $ \gamma<-6.5935 $ always has a root. For this case, the general form of the equations due to Eqs. (\ref{Ph1}) and (\ref{Ph3}) will be as follows,
 \begin{equation}\label{Ph14}
H =\frac{\sqrt{\frac{k \,r^{\frac{72 \pi +32 \gamma}{24 \pi +13 \gamma}} r +r^{3}+r -2}{r}}}{r \sin \! \left(\theta \right)}
\end{equation}
\begin{equation}\label{Ph15}
\phi^{r}=\frac{12 \left(\left(\pi +\frac{\gamma}{4}\right) r^{\frac{96 \pi +45 \gamma}{24 \pi +13 \gamma}}-2 \left(\pi +\frac{13 \gamma}{24}\right) \left(r -3\right)\right) \csc \! \left(\theta \right)}{r^{3} \left(24 \pi +13 \gamma \right)}
\end{equation}
 \begin{equation}\label{Ph16}
\phi^{\theta}=-\frac{\sqrt{\frac{k \,r^{\frac{72 \pi +32 \gamma}{24 \pi +13 \gamma}} r +r^{3}+r -2}{r}}\, \cos \! \left(\theta \right)}{r^{2} \sin \! \left(\theta \right)^{2}}
\end{equation}
According to the ranges obtained for the $\gamma$ parameter, the behavior of the model and the total topological charge can be seen in the table (\ref{4'}), which in Fig.(\ref{14}) we have drawn a sample from each area.
  \begin{figure}[H]
 \begin{center}
 \subfigure[]{
 \includegraphics[height=5cm,width=5cm]{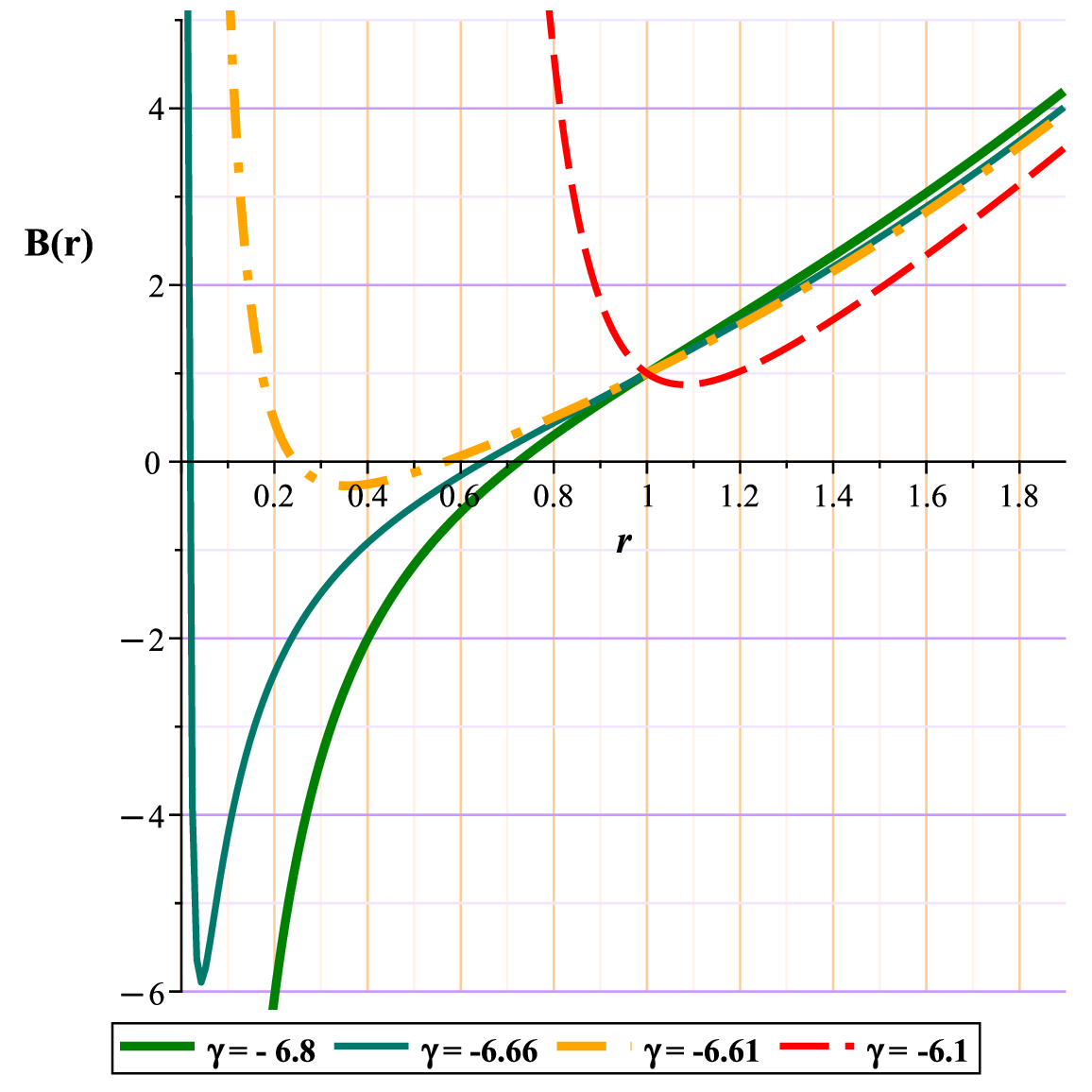}
 \label{14a}}
 \subfigure[]{
 \includegraphics[height=5cm,width=5cm]{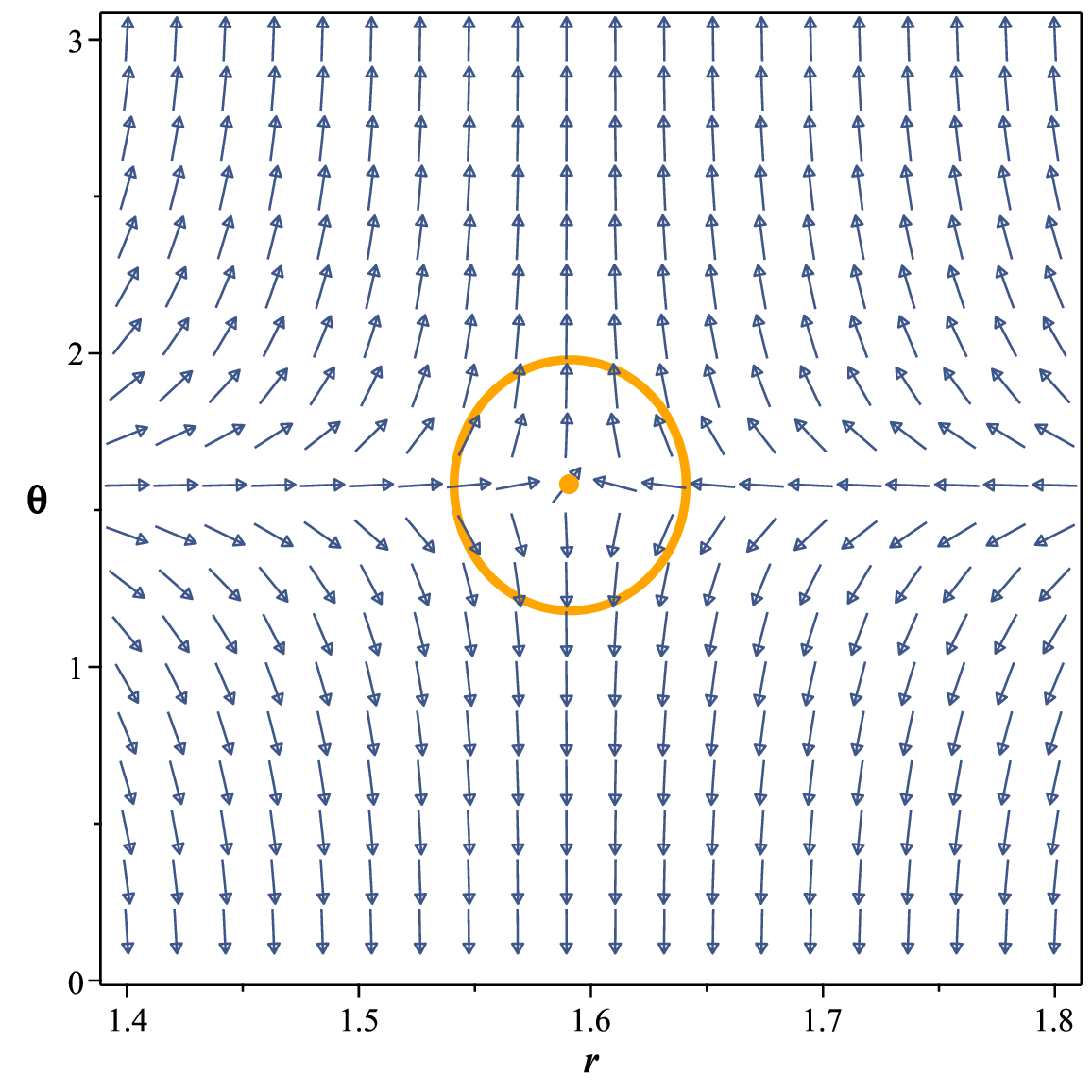}
 \label{14b}}
 \subfigure[]{
 \includegraphics[height=5cm,width=5cm]{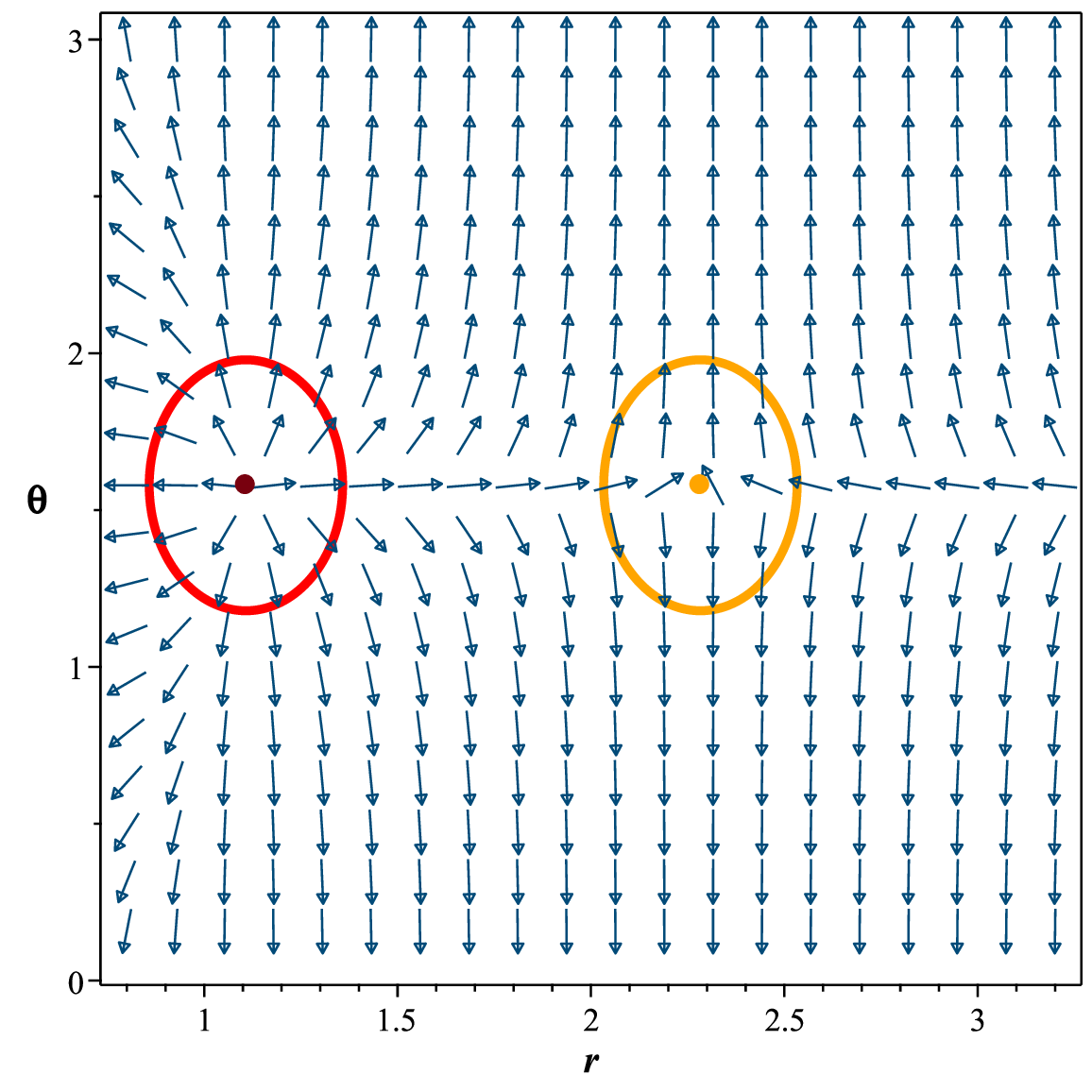}
 \label{14c}}
\caption{\small{Fig. \ref{14a}: Metric function with different $\gamma$ for Kiselev-AdS black holes in f(R, T) gravity, \ref{14b}: The normal vector field $n$ in the $(r-\theta)$ plane. The photon sphere are located at $ (r,\theta)=(1.591046,1.57)$  with respect to $( \gamma=-6.605, m=1,l=1,k=1 )$. \ref{15b}: The normal vector field $n$ in the $(r-\theta)$ plane. The photon sphere are located at $ (r,\theta)=(1.107899985366,1.57),(r,\theta)=(2.282979884819,1.57)$  with respect to $( \gamma=--6.45, m=1,l=1,k=1 )$}}
 \label{14}
\end{center}
\end{figure}
\begin{center}
\begin{table}
  \centering
\begin{tabular}{|p{3cm}|p{4cm}|p{5cm}|p{1.5cm}|p{3cm}|}
  \hline
  \centering{Kiselev-AdS Black Holes}  & \centering{Fix parametes} &\centering{Conditions}& TTC&\ $(R_{PLPS})$\\[3mm]
   \hline
  \centering{unstable photon sphere} & \centering $k=1,m=1,l=1$ & \centering{$-6.5936\leq\gamma  $} &\centering $-1$&\ $1.593135505582$ \\[3mm]
   \hline
   \centering{naked singularity} & \centering $k=1,m=1,l=1$ & \centering{$-6.5936<\gamma < -5.8$} & \centering $ 0 $ &\ $-$ \\[3mm]
   \hline
   \end{tabular}
   \caption{$R_{PLPS}$: The minimum or maximum possible radius for the appearance of an unstable photon sphere. TTC: Total Topological Charge}\label{4'}
\end{table}
 \end{center}
 Before concluding this section, it is important to note that the studies and calculations for the radiation region ($\omega = 1/3$) did not present any new insights that would lead to different conclusions compared to the three plotted regions, except for maintaining the trend of changes in the $\gamma$ parameter range. Therefore, to avoid redundancy, this region has been omitted.
\section{Conclusion}
In this paper, we have thoroughly investigated the topological charge and the conditions for the existence of the photon sphere (PS) in Kiselev-AdS black holes within \(f(R, T)\) gravity. We established their topological classifications using two distinct methods based on Duan’s topological current \(\phi\)-mapping theory viz temperature and the generalized Helmholtz free energy method. By analyzing the critical and zero points (topological charges and topological numbers) for different parameters, we revealed that the Kiselev parameter \(\omega\) and the \(f(R, T)\) gravity parameter \(\gamma\) significantly influence the number of topological charges of black holes, providing novel insights into their topological classifications. Our findings indicate that for given values of the free parameters, there exist total topological charges (\(Q_{total} = -1\)) for the T-method and total topological numbers (\(W = +1\)) for the generalized Helmholtz free energy method. Notably, in contrast to the scenario where \(\omega = 1/3\), increasing the parameter \(\gamma\) in other cases results in an increase in the number of total topological charges for the black hole. Interestingly, for the phantom field (\(\omega = -4/3\)), decreasing the parameter \(\gamma\) leads to an increase in the number of topological charges.\\

Additionally, our study of the photon sphere reveals that the simultaneous presence of \(\gamma\) and \(\omega\) effectively expands the permissible range for \(\gamma\), allowing the model to exhibit black hole behavior over a larger domain. It is evident that with the stepwise reduction of \(\omega\), the region covered by singularity diminishes and becomes more restricted. An intriguing aspect of our findings, compared to previously studied models \cite{55,56,57,58}, is the elimination of the forbidden region in this model. In other words, in the investigated areas, there is no region where both the \(\phi\) function and the metric function simultaneously lack solutions.
Also, at the end, we fully checked the curvatures singularities, and energy conditions for the mentioned black hole
In conclusion, our research provides a comprehensive understanding of the topological properties of Kiselev-AdS black holes within \(f(R, T)\) gravity. On the other hand, the influence of the parameters \(\omega\) and \(\gamma\) on the topological charges and the conditions for the existence of the photon sphere offer valuable insights into the complex behavior of these black holes. These findings pave the way for further exploration and deeper understanding of the topological aspects of black holes in modified gravity theories.\\
While our study has shed light on several important aspects of Kiselev-AdS black holes within \(f(R, T)\) gravity, it also opens up several intriguing questions for future research:\\
1.  How do these findings translate to other modified gravity theories? Can similar topological classifications be observed in other contexts, such as \(f(R)\) or \(f(T)\) gravity alone?\\
2.  What are the dynamic properties of these black holes under perturbations? How do the topological charges evolve, and what implications does this have for black hole stability?\\
3.  How do quantum corrections affect the topological properties of these black holes? Can we extend our classical findings to a quantum regime, and what new phenomena might emerge?\\
4. How do these topological properties manifest in higher-dimensional black holes? Does the dimensionality of spacetime introduce new complexities or simplifications in the topological classifications?\\
Exploring these questions will not only deepen our understanding of black holes in modified gravity theories but also potentially uncover new and exciting aspects of gravitational physics.

\end{document}